\documentclass[reprint,amsmath,amssymb,aps,pre,floatfix,superscriptaddress]{revtex4-1}

\usepackage{times,color,graphicx}
\usepackage[colorlinks=true,urlcolor=blue,linkcolor=blue,
            citecolor=blue]{hyperref}

\usepackage[export]{adjustbox}
\usepackage{placeins}
\usepackage{bm}
\usepackage{amsmath}

\newcommand{\bibtitle}[1]{\textit{#1},}
\begin{document}

\title{A tail-regression estimator for heavy-tailed distributions of
       known tail indices \\ and its application to continuum quantum
       Monte Carlo data}

\author{Pablo L\'opez R\'ios}
  \email{pl275@cam.ac.uk}
  \affiliation{Max-Planck Institute for Solid State Research,
               Heisenbergstra{\ss}e 1, 70569 Stuttgart, Germany}
  \affiliation{Theory of Condensed Matter Group, Cavendish Laboratory,
               19 J. J. Thomson Avenue, Cambridge CB3 0HE, UK}

\author{Gareth J. Conduit}
  \affiliation{Theory of Condensed Matter Group, Cavendish Laboratory,
               19 J. J. Thomson Avenue, Cambridge CB3 0HE, UK}

\begin{abstract}
Standard statistical analysis is unable to provide reliable confidence
intervals on expectation values of probability distributions that do
not satisfy the conditions of the central limit theorem.
We present a regression-based estimator of an arbitrary moment of a
probability distribution with power-law heavy tails that exploits
knowledge of the exponents of its asymptotic decay to bypass this
issue entirely.
Our method is applied to synthetic data and to energy and atomic force
data from variational and diffusion quantum Monte Carlo calculations,
whose distributions have known asymptotic forms [J. R. Trail,
\href{https://doi.org/10.1103/PhysRevE.77.016703} {Phys.\ Rev.\ E
\textbf{77}, 016703 (2008)}; A. Badinski \textit{et al.},
\href{https://doi.org/10.1088/0953-8984/22/7/074202} {J. Phys.:
Condens.\ Matter \textbf{22} 074202 (2010)}].
We obtain convergent, accurate confidence intervals on the variance of
the local energy of an electron gas and on the Hellmann-Feynman force
on an atom in the all-electron carbon dimer.
In each of these cases the uncertainty on our estimator is 45\% and 60
times smaller, respectively, than the nominal (ill-defined) standard
error.
\end{abstract}

\maketitle

%%%%%%%%%%%%%%%%%%%%%%%%%%%%%%%%%%%%%%%%%%%%%%%%%%%%%%%%%%%%%%%%%%%%%%
%%%%%%%%%%%%%%%%%%%%%%%%%%%%%%%%%%%%%%%%%%%%%%%%%%%%%%%%%%%%%%%%%%%%%%
%%%%%%%%%%%%%%%%%%%%%%%%%%%%%%%%%%%%%%%%%%%%%%%%%%%%%%%%%%%%%%%%%%%%%%
\section{Introduction}

Monte Carlo integration methods \cite{Kalos_Whitlock_montecarlo_1986}
allow the evaluation of arbitrarily complicated high-dimensional
integrals using random, discrete samples of the integrand.
Besides the need to correct for serial correlation
\cite{flyvbjerg_reblock_1989, Jonsson_reblock_2018}, the statistical
analysis of these random samples is usually straightforward, and the
final result of a Monte Carlo calculation is typically computed as a
standard mean with an accompanying standard error, defining a
confidence interval on the quantity of interest
\cite{Stroock_probability_1993}.
However there are problems for which the integrand diverges in such
manner as to render these confidence intervals invalid.

This is the case in continuum quantum Monte Carlo (QMC) methods
\cite{Ceperley_DMC_1980, Foulkes_QMC_2001}, a prominent family of
tools for studying correlated many-body systems.
Given a trial wave function $\Psi({\bf R})$, the variational Monte
Carlo (VMC) method evaluates the expectation value of an observable
$\hat A$ by accumulating its local value $A({\bf R})$ at random
real-space configurations of the particles in the system, $\{\bf R\}$,
distributed according to $\left\vert\Psi({\bf R})\right\vert^2$.
The diffusion Monte Carlo (DMC) method samples the lowest-energy wave
function $\Phi({\bf R})$ with the same nodal structure as $\Psi({\bf
R})$ by stochastic projection according to the imaginary-time
Schr\"odinger equation, and yields more accurate estimates of
observables than VMC.
The stochastic integration employed by these methods allows using
trial wave functions that are not analytically integrable, providing
extraordinary flexibility and compactness in the description of
many-body correlations \cite{LopezRios_Jastrow_2012,
LopezRios_backflow_2006, Bajdich_Pfaffians_2006}.
The VMC and DMC methods are routinely used to solve electronic
structure and quantum chemistry problems \cite{Ceperley_DMC_1980,
Seth_atoms_2011, Drummond_hydrogen_2015}.

The mismatch between the nodes of the trial wave function, $\{{\bf
R}:~\Psi({\bf R})=0\}$, and those of the true ground-state wave
function of the system are the main source of outliers in the local
energy distribution sampled in QMC, resulting in heavy tails
\cite{Panharipande_helium_1986, Trail_htail_2008} that preclude the
evaluation of meaningful confidence intervals on the estimated
variance of the local energy.
The local atomic force, comprising a Hellmann-Feynman force
\cite{Feynman_forces_1939} and a Pulay term \cite{Pulay_forces_1969,
Casalegno_forces_2003}, has heavy tails arising both from the
divergence of the electron-nucleus potential energy in all-electron
systems \cite{Badinski_forces_2010} and from the nodal error, which
prevent the evaluation of meaningful confidence intervals on the
estimated expectation value of the force.

Various methods have been proposed to circumvent the statistical
hurdles in the evaluation of atomic forces in QMC.
Modified estimators of the force satisfying a zero-variance principle
have been proposed \cite{Assaraf_zero-variance_1999,
Assaraf_forces_2001, Assaraf_forces_2003, Per_forces_2008,
Badinski_zero-variance_2008} that substantially reduce the magnitude
of the heavy tails in the local force distribution.
Force estimation methods based on the use of pseudopotentials
\cite{Badinski_pp_forces_2007, Badinski_pp_forces_2008} can eliminate
the problematic behavior of the Hellmann-Feynman force
\cite{Badinski_forces_2010}, as does the fitting approach proposed by
Chiesa \textit{et al.\@} \cite{Chiesa_force_density_2005}.
However, some of these methods involve approximations, and none of
them addresses the heavy tails in the local Pulay force distribution,
and therefore the total force remains affected by an infinite-variance
problem.
The reweighted approach proposed by Attaccalite and Sorella
\cite{Attaccalite_reweight_2008} does prevent the Pulay force from
diverging, but it involves modifying the sampling distribution and is
therefore not applicable in DMC.

Tail-index estimation methods \cite{Hill_TIE_1975, Beirlant_TIE_1996,
Kratz_qq_1996, Baek_TIE_2010} allow the exponent governing the
asymptotic behavior of heavy-tailed probability distributions to be
estimated from statistical samples.
This prior work, combined with knowledge of the exact asymptotic form
of the tails of the local energy \cite{Trail_htail_2008} and local
force \cite{Badinski_forces_2010} distributions, provides us with the
foundation to develop the \textit{tail-regression estimator} (TRE) for
heavy-tailed distributions.
We demonstrate the application of this technique to VMC and DMC data,
for which we are able to obtain forces and local energy variances not
affected by the infinite-variance problem.

The rest of this paper is structured as follows.
We review the properties of the heavy tails of the local energy and
local force distributions in section \ref{sec:htail_qmc_review}.
In section \ref{sec:mean_review} we discuss the standard method for
estimating an expectation value from a statistical sample, and we
propose our tail-regression estimator in section \ref{sec:TRE}.
The application of our methodology is illustrated in section
\ref{sec:apply_model} using model distributions of known statistical
properties.
Finally, we present the results of applying our method to the VMC
energy of a homogeneous electron gas and the VMC and DMC atomic force
on an atom in the all-electron C$_2$ molecule in sections
\ref{sec:apply_vmc} and \ref{sec:apply_dmc}, and our conclusions are
stated in section \ref{sec:conclusions}.

%%%%%%%%%%%%%%%%%%%%%%%%%%%%%%%%%%%%%%%%%%%%%%%%%%%%%%%%%%%%%%%%%%%%%%
%%%%%%%%%%%%%%%%%%%%%%%%%%%%%%%%%%%%%%%%%%%%%%%%%%%%%%%%%%%%%%%%%%%%%%
%%%%%%%%%%%%%%%%%%%%%%%%%%%%%%%%%%%%%%%%%%%%%%%%%%%%%%%%%%%%%%%%%%%%%%
\section{Heavy tails in quantum Monte Carlo}
\label{sec:htail_qmc_review}

We explore the formal definition of expectation values in QMC to allow
the characterization of the resulting heavy-tailed distributions.
For simplicity, we restrict our analysis to the VMC method until
Section \ref{sec:apply_dmc}, in which we discuss the application of
our methodology to DMC data.
Note that we ignore serial correlation in this work.

Given a trial wave function $\Psi({\bf R})$, a VMC calculation
evaluates the expectation value
\begin{equation}
  \label{eq:expval_vmc}
  \langle A\rangle =
    \frac{\int \Psi^*({\bf R}) \hat A \Psi({\bf R})
               \,{\rm d}{\bf R}}
         {\int \vert\Psi({\bf R})\vert^2
               \, {\rm d}{\bf R}} \;,
\end{equation}
by generating electronic configurations $\{\bf R\}$ according to the
distribution
\begin{equation}
  \label{eq:pr_vmc}
  P_{\bf R}({\bf R}) =
    \frac{\vert\Psi({\bf R})\vert^2}
         {\int \vert\Psi({\bf R})\vert^2 \, {\rm d}{\bf R}} \;,
\end{equation}
and evaluating the local values $A({\bf R}) = \Psi^{-1}({\bf R})
\hat A \Psi({\bf R})$ of observable $\hat A$.
The VMC expectation value can be recast as
\begin{equation}
  \label{eq:expval_integral}
  \langle A \rangle =
    \int_{-\infty}^{\infty} P_A(A) A \, {\rm d}A \;,
\end{equation}
where
\begin{equation}
  \label{eq:pa_qmc}
  P_A(A) = \int_{\partial\Omega(A)}
                \frac{P_{\bf R}({\bf R})}
                     {\vert \nabla_{\bf R}A({\bf R}) \vert}
                \,{\rm d}^{dN-1}{\bf R} \;,
\end{equation}
where $d$ is the dimensionality of the system, $N$ is the number of
electrons, and $\partial\Omega(A)$ is the $(dN-1)$-dimensional region
of configuration space where $A({\bf R})=A$.

The local value of some important observables diverge at certain
configurations, and it is often possible to characterize the
asymptotic behavior of $P_A(A)$ from knowledge of the analytical form
of $\Psi({\bf R})$ near to these configurations.
We summarize the relevant properties of the local energy and the local
atomic force below.

%%%%%%%%%%%%%%%%%%%%%%%%%%%%%%%%%%%%%%%%%%%%%%%%%%%%%%%%%%%%%%%%%%%%%%
\subsection{Local energy}
\label{sec:htail_energy}

Consider the Hamiltonian of a molecular system in Hartree atomic
units ($\hbar = m_{\rm e} = \vert e\vert = 4\pi\epsilon_0 = 1$),
\begin{equation}
  \label{eq:hamiltonian_molecule}
  \hat{H}({\bf R}) = -\frac 1 2 \sum_i \nabla_i^2 +
     \sum_{i,j>i} \frac 1 {r_{ij}} +
     \sum_{i,I} \frac {-Z_I} {r_{iI}} +
     \sum_{I,J>I} \frac {Z_I Z_J} {r_{IJ}} \;,
\end{equation}
where $r_{ij}$ is the distance between the $i$th and $j$th electrons,
$r_{iI}$ is the distance between the $i$th electron and the $I$th
nucleus, $r_{IJ}$ is the distance between the $I$th and $J$th nuclei,
and $Z_I$ is the atomic number of the $I$th nucleus.

The situations in which the local energy $E({\bf R}) = \Psi^{-1}({\bf
R}) \hat{H}({\bf R}) \Psi({\bf R})$ diverges were classified in detail
by Trail \cite{Trail_htail_2008}.
The divergence of the Coulomb potential at electron-nucleus and
electron-electron coalescence points, $r_{iI}\to 0$ and $r_{ij}\to 0$
respectively, can be neutralized by constraining the trial wave
functions to obey the Kato cusp conditions \cite{Kato_cusp_1957} under
which the kinetic energy exactly cancels the potential energy at
two-body coalescence points.
The remaining divergences of the local energy arise when $\Psi({\bf
R})\to 0$ but $\Psi$ is not locally identical to an eigenstate of
$\hat{H}$, since $\hat{H}({\bf R}) \Psi({\bf R})$ can be finite where
$\Psi({\bf R})$ is zero.

Mismatches between the nodes of $\Psi({\bf R})$ and $\hat{H}({\bf R})
\Psi({\bf R})$ are responsible for the asymptotic behavior
\cite{Trail_htail_2008}
\begin{equation}
  P_{E}(E) = c_0 \vert E-E_0\vert^{-4} + c_1 \vert E-E_0\vert^{-5}
    + \ldots \;,
\end{equation}
when $\vert E\vert\to\infty$, where $E_0$ is the exact ground state
energy and $\{c_i\}$ are unknown coefficients.
The coefficients of even powers of $\vert E-E_0\vert$ in the left
($E\to-\infty$) and right ($E\to+\infty$) tails of $P_E(E)$ have equal
coefficients, $c_{2n}^{\rm L} = c_{2n}^{\rm R}$, while those of odd
powers are of the same magnitude but of opposite signs, $c_{2n+1}^{\rm
L} = -c_{2n+1}^{\rm R}$.

The expectation value of the energy itself can be evaluated with
standard estimators without problems, but these yield unreliable
confidence intervals on the variance of the local energy.
The variance of the local energy is an important quantity since it is
directly related to the quality of the trial wave function, and is
routinely used in wave function optimization
\cite{Umrigar_varmin_1988}, as well as in the ``variance
extrapolation'' technique that attempts to estimate the zero-variance
(exact wave function) limit of expectation values
\cite{Holzmann_ueg_2003}.
We discuss the specific issues with the variance of the local energy
in section \ref{sec:mean_review}.

%%%%%%%%%%%%%%%%%%%%%%%%%%%%%%%%%%%%%%%%%%%%%%%%%%%%%%%%%%%%%%%%%%%%%%
\subsection{Local force}
\label{sec:htail_force}

The force exerted by the electrons and the other nuclei on the $I$th
nucleus of a system along the $x$ direction is $\langle
\hat{F}_{x,I}\rangle = -{\rm d}\langle \hat{H} \rangle/{\rm d}x_I$,
where $x_I$ is the $x$ Cartesian coordinate of the $I$th nucleus.
Dropping the $I$ and $x$ labels, the local force can be expressed as
\begin{equation}
  \label{eq:total_local_force}
  F({\bf R}) = F_{\rm HFT}({\bf R}) + F_{\rm P}({\bf R}) \;,
\end{equation}
where the Hellmann-Feynman force is
\begin{equation}
  \begin{split}
  F_{\rm HFT}({\bf R}) & =
    -\Psi^{-1}({\bf R})
    \frac{\partial \hat{H}({\bf R})} {\partial x}
    \Psi({\bf R}) \\ & =
    \sum_{i} \frac {-Z_I x_{iI}} {r_{iI}^3} +
    \sum_{J\neq I} \frac {Z_I Z_J x_{IJ}} {r_{IJ}^3} \;,
  \end{split}
\end{equation}
and the Pulay force is
\begin{equation}
  F_{\rm P}({\bf R}) =
    -2
    \Psi^{-1}({\bf R})
    \left[ E({\bf R}) - \langle \hat{H} \rangle \right]
    \frac{\partial \Psi({\bf R})} {\partial x} \;.
\end{equation}
Optionally, a variance-reduction term which satisfies a zero-variance
principle \cite{Assaraf_forces_2003},
\begin{equation}
  F_{\rm ZV}({\bf R}) =
    -\Psi^{-1}({\bf R})
    \left[ {\hat H}({\bf R}) - E({\bf R}) \right]
    \frac{\partial \Psi({\bf R})} {\partial x} \;,
\end{equation}
can be added to Eq.\ \ref{eq:total_local_force}.
This term does not alter the expectation value of the force but
reduces the extent of the fluctuations of the local Hellmann-Feynman
force.

The local Hellmann-Feynman force diverges at electron-nucleus
coalescence points, and its distribution exhibits a power-law tail of
the form
\begin{equation}
  P_{F_{\rm HFT}}(F) = c_0 \vert F-F_0\vert^{-5/2}
    + c_1 \vert F-F_0\vert^{-3} + \ldots \;,
\end{equation}
when $\vert F\vert\to\infty$, where $F_0$ is a constant and $\{c_i\}$
are unknown coefficients.
The coefficients of the leading-order term on the left and right tails
are equal, $c_0^{\rm L} = c_0^{\rm R}$, and the rest are asymmetric.

If the wave function satisfies the electron-nucleus Kato cusp
conditions \cite{Ma_cusp_2005} the zero-variance term exactly cancels
this divergence \cite{Per_forces_2008}, but, like for the local
energy, the mismatch between the nodes of $\Psi(\bf R)$ and
$\hat{H}({\bf R}) \Psi({\bf R})$ is responsible for the heavy tails in
the distribution of the local values of the Pulay and zero-variance
terms, and hence of the total force, which satisfies
\cite{Badinski_forces_2010}
\begin{equation}
  \label{eq:P_Ftot}
  P_{F}(F) = c_0 \vert F-F_0\vert^{-5/2}
    + c_1 \vert F-F_0\vert^{-3} + \ldots \;,
\end{equation}
when $\vert F\vert\to\infty$, where $F_0$ is a constant and $\{c_i\}$
are unknown coefficients exhibiting no symmetry.

A somewhat different scenario arises if nondivergent pseudopotentials
are used in place of the electron-nucleus Coulomb potential
\cite{Trail_pp1_2015, Trail_pp2_2017} and the local force estimation
is consequently adjusted \cite{Badinski_pp_forces_2007,
Badinski_pp_forces_2008}.
In this case the local Hellmann-Feynman force exhibits less
problematic heavy tails of leading order $\vert F-F_0 \vert^{-4}$
\cite{Badinski_forces_2010}, but since the Pulay term is unaffected by
the use of pseudopotentials the local total force remains of the form
of Eq.\ \ref{eq:P_Ftot}.

%%%%%%%%%%%%%%%%%%%%%%%%%%%%%%%%%%%%%%%%%%%%%%%%%%%%%%%%%%%%%%%%%%%%%%
%%%%%%%%%%%%%%%%%%%%%%%%%%%%%%%%%%%%%%%%%%%%%%%%%%%%%%%%%%%%%%%%%%%%%%
%%%%%%%%%%%%%%%%%%%%%%%%%%%%%%%%%%%%%%%%%%%%%%%%%%%%%%%%%%%%%%%%%%%%%
\section{Standard estimation of an expectation value}
\label{sec:mean_review}

The expectation value of an observable whose local value $A$ is
distributed according to $P_A(A)$ is
\begin{equation}
  \label{eq:expval_integral}
  \langle A \rangle =
    \int_{-\infty}^{\infty} P_A(A) A \, {\rm d}A \;,
\end{equation}
and the variance of $A$ is the expectation value of $\left(A - \langle
A\rangle \right)^2$,
\begin{equation}
  \label{eq:expval_variance}
  {\rm Var}[A] = \sigma^2 =
    \int_{-\infty}^{\infty}
      P_A(A) \left(A-\langle A \rangle\right)^2 \, {\rm d}A \;.
\end{equation}
The integrals in Eqs.\ \ref{eq:expval_integral} and
\ref{eq:expval_variance} must be nondivergent for the expectation
value and variance of $A$ to be well-defined.
Therefore, probability distributions with asymptotic behavior $P_A(A)
\sim \vert A\vert^{-\mu}$ as $\vert A\vert\to\infty$ have no
well-defined expectation value or variance for $\mu\leq 2$, and have a
well-defined expectation value but no well-defined variance for
$2<\mu\leq 3$.
Note that a function with $\mu\leq 1$ is not a valid probability
distribution as it cannot be normalized.

Let $\{A_m\}_{m=1}^M$ be a sample of $M$ independent random variables
identically distributed according to $P_A(A)$.
The standard estimator for Eq.\@ \ref{eq:expval_integral} is the
sample mean,
\begin{equation}
  \bar A = \frac 1 M \sum_{m=1}^M A_m \;,
\end{equation}
and the standard estimator for Eq.\@ \ref{eq:expval_variance} is the
sample variance,
\begin{equation}
  S^2 = \frac {\sum_{m=1}^M (A_m-{\bar A})^2}
                {M-1} \;.
\end{equation}
The uncertainty on $\bar A$ is the standard error $\sigma_{\bar A} =
S/\sqrt{M}$.
This poses a problem for distributions of leading-order exponent
$2<\mu\leq 3$:\@ despite having a well-defined expectation value
according to Eq.\ \ref{eq:expval_integral}, its standard estimator has
a divergent uncertainty because it is defined in terms of the
divergent variance of Eq.\ \ref{eq:expval_variance}.

In this regime $P_A(A)$ does not satisfy the conditions of the central
limit theorem that would guarantee the asymptotic normality of
confidence intervals built from the standard mean and standard error
\cite{Stroock_probability_1993}.
Instead, $P_A(A)$ satisfies the law of large numbers, which states
that the standard mean does converge to the expectation value at
infinite sample size but confidence intervals cannot be constructed
using the standard error as finite sample sizes.

A similar issue affects the estimator of the variance itself.
Even though the variance is well-defined for $\mu>3$, the variance
on the estimator of the variance is
\begin{equation}
  {\rm Var}[S^2] = \frac 1 M
                   \left(
                     m_4 - \frac{M-3}{M-1}\sigma^4
                   \right) \;,
\end{equation}
where $m_4$ is the fourth-order central moment of $P_A(A)$, which
diverges for distributions of leading-order exponent $\mu\leq 5$,
leading to a divergent uncertainty on the $S^2$ estimator of the
variance.

%%%%%%%%%%%%%%%%%%%%%%%%%%%%%%%%%%%%%%%%%%%%%%%%%%%%%%%%%%%%%%%%%%%%%%
%%%%%%%%%%%%%%%%%%%%%%%%%%%%%%%%%%%%%%%%%%%%%%%%%%%%%%%%%%%%%%%%%%%%%%
%%%%%%%%%%%%%%%%%%%%%%%%%%%%%%%%%%%%%%%%%%%%%%%%%%%%%%%%%%%%%%%%%%%%%
\section{Tail-regression estimator}
\label{sec:TRE}

As outlined in the previous section, the uncertainty on the standard
estimator of a moment of a probability distribution involves the
estimator of a higher-order moment, which might be divergent even
though the moment of interest is well-defined.
We therefore propose an alternative estimator of the moment of a
heavy-tailed probability distribution that exploits knowledge of its
analytical asymptotic form to yield well-defined confidence intervals
whenever the moment itself is well defined.

Without loss of generality we focus on distributions with a right
heavy tail; the extension of our analysis to distributions with left
and both left and right heavy tails is straightforward.
In particular, we consider a probability distribution exhibiting a
right tail of asymptotic form
\begin{equation}
  \label{eq:tail_form_asymptotic}
  P_A(A) = \sum_n c_n \left|A-A_0\right|^{-\mu-n\Delta} \;,
\end{equation}
when $A\to\infty$, where the leading-order exponent $\mu$ and exponent
increment $\Delta>0$ are assumed to be known analytically, as is the
case for local energies, with $\mu=4$ and $\Delta=1$
\cite{Trail_htail_2008}, and for local forces, with $\mu=5/2$ and
$\Delta=1/2$ \cite{Badinski_thesis_2008}.
The specific value of $\Delta$ is not critical for the correct
description of the asymptote by Eq.\ \ref{eq:tail_form_asymptotic},
and can be assumed to be unity, but the accuracy of a truncated
expansion strongly depends on $\Delta$.
In other words, the bias incurred by choosing a suboptimal value of
$\Delta$ can be made arbitrarily small by using a larger expansion.
In Eq.\ \ref{eq:tail_form_asymptotic} $\{c_n\}$ are unknown
coefficients, and $A_0$ is an unknown parameter which is assumed to
lie close to the ``center'' of the distribution.

%%%%%%%%%%%%%%%%%%%%%%%%%%%%%%%%%%%%%%%%%%%%%%%%%%%%%%%%%%%%%%%%%%%%%%
\subsection{Validity of asymptote with approximate $A_0$}
\label{sec:A0_Ac}

First, we address the fact that $A_0$ is an unknown nonlinear
parameter in Eq.\ \ref{eq:tail_form_asymptotic} and will have to be
approximated.
Let $A_{\rm c}$ be an approximation to $A_0$ such that $A_0=A_{\rm
c}+\varepsilon$, where $\varepsilon$ is a small error.
Assuming for simplicity that $\Delta^{-1}$ is an integer and
expanding to first order in $\varepsilon$ we find
\begin{equation}
  \label{eq:tail_form_Ac_error}
  \begin{split}
  P_A(A) = & \sum_n c_n |A-A_{\rm c}-\varepsilon|^{-\mu-n\Delta} \\
       \approx & \sum_n c_n |A-A_{\rm c}|^{-\mu-n\Delta} \\
       & + \varepsilon\sum_n c_n  (\mu+n\Delta)
                  |A-A_{\rm c}|^{-\mu-n\Delta-1} \\
       = & \sum_n c_n^\prime |A-A_{\rm c}|^{-\mu-n\Delta}
       = P_A^\prime(A) \;, \\
  \end{split}
\end{equation}
where
\begin{equation}
\label{eq:prime_params}
c_n^\prime = \left\{
  \begin{array}{l@{\,,\quad}l}
    c_n                                                &
       n<\Delta^{-1}    \\
    c_n + \varepsilon c_{n-\Delta^{-1}}(\mu+n\Delta-1) &
       n\geq\Delta^{-1} \\
  \end{array}
 \right. \;.
\end{equation}
The asymptotic expression $P_A^\prime(A)$ has the same form as
$P_A(A)$, albeit with modified coefficients $c_n^\prime$ for $n\geq
\Delta^{-1}$.
We shall therefore proceed with the derivation of our estimator using
$P_A^\prime(A)$ as the asymptote, dropping the primes from the
notation for clarity.
This effectively amounts to replacing $A_0$ with $A_{\rm c}$, which in
practice we set to the sample median.

%%%%%%%%%%%%%%%%%%%%%%%%%%%%%%%%%%%%%%%%%%%%%%%%%%%%%%%%%%%%%%%%%%%%%%
\subsection{Estimator}

In order to develop our estimator, we start by assuming that there
exists a threshold $A_{\rm R}$ such that for $A>A_{\rm R}$ the
probability distribution is accurately represented by an expansion of
order $n_{\rm R}$,
\begin{equation}
  \label{eq:tail_form}
  P_A(A) = \sum_{n=0}^{n_{\rm R}} c_n
           \left|A-A_{\rm c}\right|^{-\mu-n\Delta}
           \,,\quad A>A_{\rm R} \;.
\end{equation}
The integral of Eq.\ \ref{eq:expval_integral} can be partitioned at
$A_{\rm R}$ into central and right-tail contributions,
\begin{equation}
  \label{eq:expval_split}
  \langle A \rangle =
    \int_{-\infty}^{A_{\rm R}} P_A(A) A \, {\rm d}A
      + \int_{A_{\rm R}}^{\infty} P_A(A) A \, {\rm d}A \;.
\end{equation}
Let $\{A_m\}$ be a sample of $M$ independent random variables
identically distributed according to $P_A(A)$, $\{A^{(m)}\}$ the
corresponding order statistics, \textit{i.e.}, the re-indexed version
of $\{A_m\}$ such that $A^{(1)}>A^{(2)}>\ldots>A^{(M)}$, $M_{\rm C}$
the number of data in the central region $A<A_{\rm R}$, and $M_{\rm
R}=M-M_{\rm C}$ the number of data in the tail.
We define the \textit{tail-regression estimator} of $\langle A\rangle$
as
\begin{equation}
  \label{eq:TRE}
  {\cal A} =
    \frac 1 M \sum_{m>M_{\rm R}} A^{(m)} +
    \sum_{n=0}^{n_{\rm R}} c_n \int_{A_{\rm R}}^{\infty}
    \left|A-A_{\rm c}\right|^{-\mu-n\Delta}
    A \, {\rm d}A \;.
\end{equation}
The integrals in Eq.\ \ref{eq:TRE} are nondivergent for $\mu>2$ and
can be evaluated analytically,
\begin{equation}
  \label{eq:mom1_integral}
  \begin{gathered}
    \int_{A_{\rm R}}^{\infty}
         \vert A-A_{\rm c} \vert^{-\mu-n\Delta}
         A \, {\rm d}A = \\
    = \left[
        \frac{\vert A_{\rm R}-A_{\rm c} \vert^{-\mu-n\Delta+2}}
             {\mu+n\Delta-2}
        + A_{\rm c}
        \frac{\vert A_{\rm R}-A_{\rm c} \vert^{-\mu-n\Delta+1}}
             {\mu+n\Delta-1}
      \right] \;.
  \end{gathered}
\end{equation}
The parameters $\{c_n\}$ in Eq.\ \ref{eq:TRE} can be obtained by
regression of $\{A^{(m)}\}_{m=1}^{M_{\rm R}}$ to Eq.\
\ref{eq:tail_form}.
Since $\cal A$ is linear in $\{c_n\}$, the distribution of $\cal A$
follows that of $\{c_n\}$.
This implies that if the regression coefficients are asymptotically
normally distributed, ${\cal A}$ will also be asymptotically normally
distributed.
We will address the distribution of regression coefficients in section
\ref{sec:tail_regression}.
We regard $A_{\rm R}$, and $n_{\rm R}$ as external parameters that we
deal with separately, see section \ref{sec:nparam_anchor}, and do not
contribute to the uncertainty on ${\cal A}$.

Analogously, we define the tail-regression estimators of the norm,
\begin{equation}
  \label{eq:TRE_mom0}
  {\cal W} =
    \frac{M_{\rm C}} M +
    \sum_{n=0}^{n_{\rm R}} c_n \int_{A_{\rm R}}^{\infty}
    \left|A-A_{\rm c}\right|^{-\mu-n\Delta} \, {\rm d}A \;,
\end{equation}
and of the variance of the distribution,
\begin{equation}
  \label{eq:TRE_mom2}
  \begin{split}
  {\cal V} & =
    \frac 1 {M-1} \sum_{m>M_{\rm R}} (A^{(m)}-{\cal A})^2 \\ & +
    \sum_{n=0}^{n_{\rm R}} c_n \int_{A_{\rm R}}^{\infty}
    \left|A-A_{\rm c}\right|^{-\mu-n\Delta}
    (A-{\cal A})^2\, {\rm d}A \;.
  \end{split}
\end{equation}
These integrals can likewise be evaluated analytically.
Finally, we note that the threshold $A_{\rm R}$ must lie on the
midpoint between two adjacent sample points, $\frac 1 2
\left(A^{(M_{\rm R})} + A^{(M_{\rm R}+1)} \right)$, to ensure that the
central contributions have the correct weight in Eqs.\ \ref{eq:TRE},
\ref{eq:TRE_mom0}, and \ref{eq:TRE_mom2}.

We use the bootstrap method \cite{Efron_bootstrap_1986} to compute
the uncertainty on ${\cal A}$.
We generate $n_{\rm bs}$ resamples of $\{A_m\}$ with replacement, that
is, the $i$th resample $\{A_{m}^{[i]}\}$ contains $M$ elements from
$\{A_m\}$ drawn at random and uniformly, allowing repetitions.
For each resample we evaluate ${\cal A}^{[i]}$, and we evaluate the
uncertainty on ${\cal A}$ as the standard deviation of the values
of $\{{\cal A}^{[i]}\}$.
Estimates of other statistical parameters arising from analysis of
$\{A_m\}$, including $\cal W$ and $\cal V$, are also obtained in this
process.
In our applications of the tail-regression estimator we use $n_{\rm
bs}=4096$ bootstrap resamples, which provide a $1.1\%$ uncertainty on
the estimated uncertainties assuming normality.
The computational cost of this approach is proportional to $M\times
n_{\rm bs}$.

%%%%%%%%%%%%%%%%%%%%%%%%%%%%%%%%%%%%%%%%%%%%%%%%%%%%%%%%%%%%%%%%%%%%%%
\subsection{Tail form in $yx$ scale}
\label{sec:tail_form_yx}

The framework for our tail-regression procedure is inspired by
regression-based tail-index estimation methods \cite{Hill_TIE_1975,
Beirlant_TIE_1996}, discussed in Appendix \ref{sec:TIE}.
First, note that the complementary cumulative distribution function
$\bar F_A(A)$ associated with $P_A(A)$ should be approximately equal
to the sample quantiles $q_m$,
\begin{equation}
  \label{eq:quantile_relation}
  \bar F_A(A) =
  \int_{A^{(m)}}^\infty P_A(A) dA \approx q_m \;,
\end{equation}
for which we use the symmetric form $q_m=\frac{m-1/2}M$.
Substituting Eq.\ \ref{eq:tail_form} into Eq.\
\ref{eq:quantile_relation} yields
\begin{equation}
  \label{eq:tail_fit}
  \sum_{n=0}^{n_{\rm R}} \frac{c_n}{\mu+n\Delta-1}
         \left|A^{(m)}-A_{\rm c}\right|^{-\mu-n\Delta+1} = q_m \;,
\end{equation}
which we rearrange as
\begin{equation}
  \label{eq:fit_function}
  q_m \left|A^{(m)}-A_{\rm c}\right|^{\mu-1} =
    \sum_{n=0}^{n_{\rm R}} \frac{c_n}{\mu+n\Delta-1}
    \left| A^{(m)}-A_{\rm c} \right|^{-n\Delta} \;.
\end{equation}
We define
\begin{equation}
  \label{eq:yx_def}
  \begin{split}
    y_m & = q_{m} \left|A^{(m)}-A_{\rm c}\right|^{\mu-1} \;, \\
    x_m & = |A^{(m)}-A_{\rm c}|^{-\Delta}  \;,
  \end{split}
\end{equation}
which we refer to as ``$yx$ scale'', under which Eq.\
\ref{eq:fit_function} reads
\begin{equation}
  \label{eq:fit_function_yx}
  y_m=\sum_{n=0}^{n_{\rm R}} \frac{c_n}{\mu+n\Delta-1} x_m^n \;,
\end{equation}
that is, $y$ is simply a polynomial in $x$.
By construction, $y$ must be positive and tend to a finite value as
$x\to 0$, $y(0)=c_0/(\mu-1)$, and the $n$th derivative of $y(x)$ at
$x=0$ is likewise proportional to $c_n$.

\begin{figure}[htb!]
  \centering
  \includegraphics[width=0.47\textwidth,valign=b]{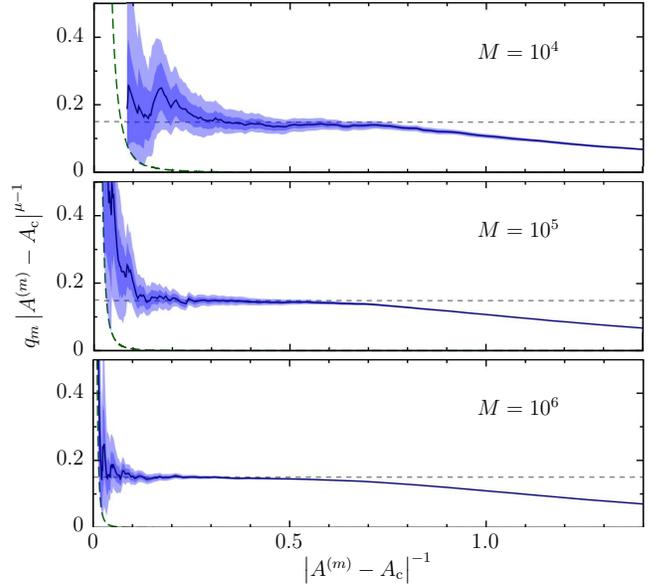}
  \caption{
    $yx$-scale plot of the left tail of $M$ independent random
    numbers identically distributed according to $H_4(A)$ for sample
    sizes $M=10^4$ (top), $10^5$ (middle), and $10^6$ (bottom).
    Shaded areas around curves correspond to the $68.3\%$ and $95.4\%$
    confidence intervals obtained from the bootstrap.
    The short-dashed lines indicate the analytical value of $y(0)$,
    and the long-dashed lines are the first-quantile lines,
    $y=\frac{1/2}{M}x^{-3}$, marking the region where the largest
    value of $A$ falls at each sample size.
    \label{fig:yx_converge_demo}}
\end{figure}

It is useful to inspect the basic properties of the $yx$ scale we
have introduced.
For illustration purposes, let
\begin{equation}
  \label{eq:ptest}
  H_\mu(A) =
    \frac{\mu \sin\frac\pi\mu}{2\pi}
    \frac 1 {1+\left\vert A\right\vert^\mu} \;,
\end{equation}
which for $\mu>2$ is a normalized probability distribution whose
expectation value is zero and has the asymptote $\vert A\vert^{-\mu}$
as $\vert A\vert\to\infty$.
In Fig.\ \ref{fig:yx_converge_demo} we show a $yx$-scale plot of the
left tail of $M$ independent random numbers identically distributed
according to $H_4(A)$ at different sample sizes $M$, assuming
$\Delta=1$.
The exact value $y(0)=0.1501$ is shown as a short-dashed line in each
panel.
The first quantile $q_1=1/(2M)$, corresponding to the largest value of
$A$ in the sample, gets smaller as $M$ increases, and, as follows from
Eq.\ \ref{eq:yx_def}, the extreme point $(x_1,y_1)$ satisfies
$y_1=q_1 x_1^{(1-\mu)/\Delta}$.
This curve determines how far left the plot extends, as shown by the
long-dashed line in each of the panels of Fig.\
\ref{fig:yx_converge_demo}.
With increasing $M$ the plot is populated from right to left,
gradually producing a better resolved curve near $x=0$.

\begin{figure}[htb!]
  \centering
  \includegraphics[width=0.47\textwidth,valign=b]{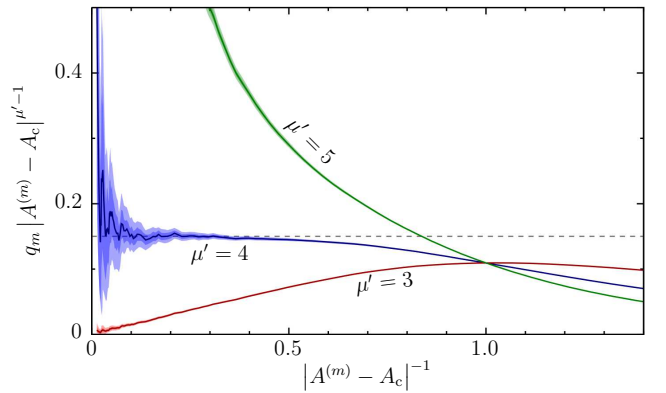}
  \caption{
    $yx$-scale plot of the left tail of $10^6$ independent random
    numbers identically distributed according to $H_4(A)$ using
    the correct leading-order tail exponent $\mu^\prime=\mu=4$ and
    incorrect values $\mu^\prime=3$ and $5$, which respectively go to
    zero and diverge as $x\to 0$.
    Shaded areas around curves correspond to the $68.3\%$ and $95.4\%$
    confidence intervals obtained from the bootstrap.
    The short-dashed line indicates the analytical value of $y(0)$.
    \label{fig:yx_demo}}
\end{figure}

In Fig.\ \ref{fig:yx_demo} we show $yx$ plots of the $M=10^6$ sample
used in Fig.\ \ref{fig:yx_converge_demo} in which the $yx$ scale is
defined using the exponents $\mu^\prime=3$, $4$, and $5$.
If the true exponent is underestimated, $\mu^\prime<\mu$, the curve
goes to zero as $x\to 0$, while overestimating the exponent,
$\mu^\prime>\mu$, yields a diverging curve at $x\to 0$.
Attempting to use $yx$-scale plots to estimate $\mu$ is rudimentary at
best, since each possible value must be tested explicitly, and care
must be taken in interpreting the large statistical fluctuations of
$y$ at small $x$.
Tail-index estimation methods \cite{Beirlant_TIE_1996, Kratz_qq_1996}
offer a robust alternative; the regression method of Ref.\
\cite{Beirlant_TIE_1996} produces a reasonable value of $\mu=3.94(2)$
from the data plotted in Fig.\ \ref{fig:yx_demo}.
Note that the tail-regression estimator relies on analytical knowledge
of $\mu$ rather than on its estimation from the sample.

%%%%%%%%%%%%%%%%%%%%%%%%%%%%%%%%%%%%%%%%%%%%%%%%%%%%%%%%%%%%%%%%%%%%%%
\subsection{Tail regression and weights}
\label{sec:tail_regression}

We now focus on the regression procedure to ensure a reliable
estimation of the fit parameters.
It is apparent from Fig.\ \ref{fig:yx_converge_demo} that the
distribution of $y$ values at small $x$ about the exact $y(0)$ is
skewed towards large $y$, so adequate least-squares weights are needed
to ensure the faithfulness of the resulting fit.

Let $r_m = w_m[y_m-y(x_m)]$ be the least-square residuals, where $w_m$
is the least-squares weight applied to the $m$th point, and
$\chi^2=\frac 1 {M-n_{\rm R}-1} \sum_{m=1}^M r_m^2$ be the
least-squares function.
Minimizing $\chi^2$ with respect to the fit parameters $\{c_n\}$
yields the linear system of equations $T {\bf c} = {\bf b}$,
where
\begin{equation}
  \label{eq:lsq_matrices}
  \begin{split}
  (T)_{pq} & = \frac 1 M \sum_{m=1}^M w_m x_m^{p+q} \;, \\
  ({\bf b})_p & = \frac 1 M \sum_{m=1}^{M} w_m y_m x_m^p \;, \\
  ({\bf c})_p & = c_p \;, \\
  \end{split}
\end{equation}
and indices $p$ and $q$ run between $0$ and $n_{\rm R}$.
The parameter vector is thus
\begin{equation}
\label{eq:coeff_vector}
{\bf c}
  = T^{-1} {\bf b}
  = \frac 1 {\det(T)} {\rm adj}(T) {\bf b} \;,
\end{equation}
where ${\rm adj}(T)$ is the adjugate matrix of $T$ (\textit{i.e.}, the
transpose of its cofactor matrix) and $\det(T)$ is its determinant.

The parameters $({\bf c})_p$ will be asymptotically normally
distributed if $(T)_{pq}$, $({\bf b})_p$, and $\det(T)$ are themselves
asymptotically normally distributed and $\det(T)$ is nonzero, since
Eq.\ \ref{eq:coeff_vector} involves sums and multiplications, which
preserve asymptotic normality, and a division, which also preserves
asymptotic normality if the denominator is strictly nonzero.

We consider weights of the form
\begin{equation}
  \label{w_form}
  w_m = |A^{(m)}-A_{\rm c}|^{-\gamma (\mu-1)} \;,
\end{equation}
where $\gamma$ is a positive constant.
Since $x_m$ is a negative power of $|A^{(m)}-A_{\rm c}|$, all elements
of $T$ and $\det(T)$ are asymptotically normally distributed, but the
elements of $\bf b$ need not be.
We focus on the $m=1$ contribution to $({\bf b})_0$, which is the
least likely to exhibit asymptotic normality,
\begin{equation}
  w_1 y_1 = q_1 \left|A^{(1)}-A_{\rm c}\right|^{(\mu-1)(1-\gamma)} \;.
\end{equation}
Noting that $w_1=q_1^\gamma y_1^{-\gamma}$ and that the exact
asymptotic value of $y_1$ as $M\to\infty$ is by construction
$y(0)=c_0/(\mu-1)$, the asymptotic limit of $w_1 y_1$ is
\begin{equation}
  \xi = q_1^{\gamma} \left[ \frac{c_0}{\mu-1} \right]^{1-\gamma} \;,
\end{equation}
where $q_1=1/(2M)$ encodes the sample-size dependence.

We now investigate the distribution of values of $w_1 y_1$ about
$\xi$.
The probability that $|A^{(1)}-A_{\rm c}|$ is bounded from above by
$\alpha$ is the probability that the $M$ points in the sample are
bounded from above by $A_{\rm c}+\alpha$, that is,
\begin{equation}
  \label{eq:cprob_extreme}
  \begin{split}
  {\rm Prob}(|A^{(1)}-A_{\rm c}|\leq \alpha)
    = & 1- \bar F_{|A^{(1)}-A_{\rm c}|}(\alpha) \\
    = & \left[1-\bar F_{A}(A_{\rm c}+\alpha)\right]^M \\
    \approx & \left(1-c_0\alpha^{-\mu+1}\right)^M \;,
  \end{split}
\end{equation}
to leading order for large $\alpha$.
Differentiating with respect to $\alpha$ yields the probability
distribution of the extreme value,
\begin{equation}
  \label{eq:prob_extreme}
  P_{\vert A^{(1)}-A_{\rm c}\vert}(\alpha)
    = M c_0 (\mu-1) \alpha^{-\mu}
      \left(1-c_0\alpha^{-\mu+1}\right)^{M-1} \;,
\end{equation}
and by a change of variable,
\begin{equation}
  \label{eq:prob_extreme_w1y1_xi}
  P_{w_1 y_1 / \xi}(\beta) =
    \frac{\mu-1}{2(1-\gamma)}
    \beta^{-\frac{2-\gamma}{1-\gamma}}
    \left[ 1 - \frac{\mu-1}{2M} \beta^{-\frac 1{1-\gamma}}
    \right]^{M-1} \;,
\end{equation}
assuming $\gamma\neq 1$.
At large $\beta$,
\begin{equation}
  \label{eq:prob_extreme_w1y1_xi_leading}
  P_{w_1 y_1 / \xi}(\beta) \sim \beta^{-\frac{2-\gamma}{1-\gamma}} \;,
\end{equation}
which is a power-law tail for $\gamma<1$, yielding an undefined
expectation value for $\gamma\leq 1/2$.
Unweighted fits ($\gamma=0$) are therefore numerically ill-conditioned
since the residuals are themselves heavy tailed $\sim\beta^{-2}$
regardless of the value of $\mu$.
Fit weights with $\gamma>1/2$ must therefore be used.

\begin{figure}[htb!]
  \centering
  \includegraphics[width=0.47\textwidth,valign=b]{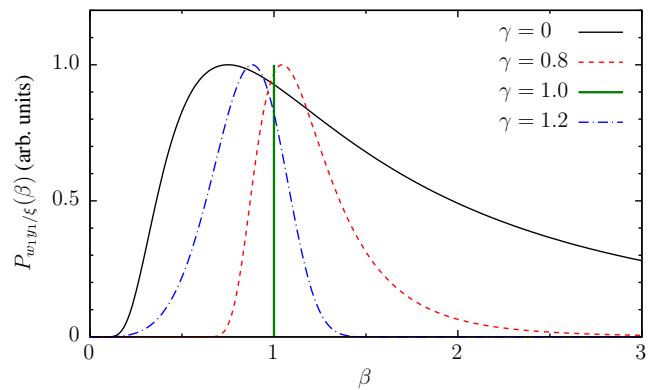}
  \caption{
    Probability distribution of the values of $w_1 y_1 / \xi$,
    Eq.\ \ref{eq:prob_extreme_w1y1_xi}, for various values of
    $\gamma$, using $\mu=4$ and $M=10^6$.
    \label{fig:wy_distrib}}
\end{figure}
In Fig.\ \ref{fig:wy_distrib} we plot $P_{w_1 y_1 / \xi}(\beta)$ for
various values of $\gamma$.
Note that the curve for $\gamma=0$ describes the distribution of
values of $y$ in Fig.\ \ref{fig:yx_converge_demo} relative to the
analytical value of $y(0)$ along the constant-$q$ lines.
The plots for $\gamma=0.8$ and $1.2$ reveal that, while not
ill-conditioned, the significant skewness of the distributions makes
the first moment of $P_{w_1 y_1 / \xi}(\beta)$ differ significantly
from the asymptotic expectation value of $1$ in these cases.
For $\gamma=1$, $w_1 y_1 = q_1^{\gamma}$ is $y_1$-independent, and the
distribution is therefore a delta function peaked at the asymptotic
expectation value, as shown in Fig.\ \ref{fig:wy_distrib}.
We therefore use $\gamma=1$ for our fit weights.

We empirically find it advantageous to include a $q_m$-dependent
factor in the fit weights so as to ensure the continuity of the fit to
the probability distribution near $A_{\rm R}$.
We choose this factor to be the weights corresponding to the
formulation of the Hill estimator of the first-order tail index
\cite{Hill_TIE_1975} as a regression estimator
\cite{Beirlant_TIE_1996}, see Eq.\ \ref{eq:beirlant_weight} in
Appendix \ref{sec:TIE}.
Therefore, our full fit weights are
\begin{equation}
  \label{eq:fit_weights}
  w_m = \left( \log\frac{q_{M_{\rm R}+1}}{q_m} \right)^{-1}
        |A^{(m)}-A_{\rm c}|^{-\mu+1} \;.
\end{equation}

%%%%%%%%%%%%%%%%%%%%%%%%%%%%%%%%%%%%%%%%%%%%%%%%%%%%%%%%%%%%%%%%%%%%%%
\subsection{Selecting $n_{\rm R}$ and $A_{\rm R}$}
\label{sec:nparam_anchor}

The tail-regression estimator depends parametrically on the expansion
order $n_{\rm R}$ and threshold $A_{\rm R}$.
In our tests we try several thresholds and converge the fit with
respect to the expansion order at each of them.
We then choose the value of $A_{\rm R}$ which minimizes the
uncertainty on either $\cal V$, if well-defined, or on $\cal A$.

We choose the expansion order heuristically by finding plateaus in
$\cal W$, $\cal A$, $\cal V$, and $\chi^2$ as a function of $n_{\rm
R}$, and selecting the smallest expansion order at which these four
functions have converged.
To ensure correctness, we further require that ${\cal W} \approx 1$
and $y(x)>0$ within the fit range, and we restrict $n_{\rm R}\geq
1/\Delta$ in order to ``absorb'' the error incurred by approximating
$A_0$ by $A_{\rm c}$, as explained in Section \ref{sec:A0_Ac}.

Note that we do not set $A_{\rm R}$ directly, but instead set $q_{\rm
R}=(M_{\rm R}-1/2)/M$, as keeping the number of sample points in each
partition fixed across bootstrap resamples eliminates the variation of
the central contribution to $\cal W$, which is statistically
advantageous.
We choose our values of $q_{\rm R}$ using a grid of equally-spaced
values of $-\log q_{\rm R}$.
We pick $n_{\rm R}$ and $A_{\rm R}$ using $n_{\rm bs}=256$, and
evaluate the final result separately with $n_{\rm bs}=4096$ to avoid
selection bias.
We illustrate the procedure for choosing $n_{\rm R}$ and $A_{\rm R}$
in section \ref{sec:model_mom4}.

%%%%%%%%%%%%%%%%%%%%%%%%%%%%%%%%%%%%%%%%%%%%%%%%%%%%%%%%%%%%%%%%%%%%%%
\subsection{Two-tailed distributions, symmetry, and constraints}
\label{sec:constraints}

The tail-regression estimator described so far can be modified
trivially for distributions with left and right heavy tails.
For simplicity we use the same expansion orders and thresholds on both
tails, $n_{\rm R}=n_{\rm L}$ and $M_{\rm R}=M_{\rm L}$.

In some important cases, including the local energy and local atomic
force in VMC \cite{Trail_htail_2008, Badinski_forces_2010}, the
leading-order coefficients of the left and right tails, $c_0^{\rm L}$
and $c_0^{\rm R}$, are equal.
This can be exploited by unifying the regression step for both tails
and imposing the constraint $c_0^{\rm L} = c_0^{\rm R} = c_0$.
The leading order contribution to $\cal A$ from the tails is
\begin{equation}
  \label{eq:symmetric_mom1_integral}
  c_0 \int_{A_{\rm R}}^{2 A_{\rm c}-A_{\rm L}}
  \left|A-A_{\rm c}\right|^{-\mu}
  A \, {\rm d}A
  \;.
\end{equation}
The exact cancellation of part of the left- and right-tail
contributions to $\cal A$ should provide a substantial reduction to
its uncertainty.
The effect on the uncertainty on $\cal V$ of enforcing symmetry can be
expected to be marginal since both tails contribute positively in this
case.

As implied by Eq.\ \ref{eq:prime_params}, due to the approximation
$A_0\approx A_{\rm c}$ constraints must not be applied to parameters
$c_n$ with $n\geq 1/\Delta$.
For example, even if a distribution with $\Delta=1$ is known to
analytically satisfy $c_1^{\rm L} = c_1^{\rm R}$, the values of the
$c_1$ parameters on each tail must be allowed to differ to account for
the error in $A_{\rm c}$.
In the tests carried out in our present work we use at most one
constraint, and we use the labels ``TRE'' and ``TRE(1)'' in the plots
in sections \ref{sec:apply_model}, \ref{sec:apply_vmc}, and
\ref{sec:apply_dmc} to distinguish the unconstrained and constrained
estimators.

The use of constraints allows for the interesting possibility of
estimating the expectation value of distributions with $1<\mu\leq 2$
for which $\langle A\rangle$ is formally divergent.
In this case we redefine the expectation value as the Cauchy principal
value of the integral in Eq.\ \ref{eq:expval_integral} with respect to
$A_0$,
\begin{equation}
  \label{eq:expval_integral_cauchy}
  \langle A \rangle =
    \lim_{a\to\infty}
      \int_{A_0-a}^{A_0+a}
        P_A(A) A \, {\rm d}A \;,
\end{equation}
ensuring that the divergent leading-order contributions cancel out due
to symmetry.
We present an example of this in section \ref{sec:model_mom1}.

%%%%%%%%%%%%%%%%%%%%%%%%%%%%%%%%%%%%%%%%%%%%%%%%%%%%%%%%%%%%%%%%%%%%%%
%%%%%%%%%%%%%%%%%%%%%%%%%%%%%%%%%%%%%%%%%%%%%%%%%%%%%%%%%%%%%%%%%%%%%%
%%%%%%%%%%%%%%%%%%%%%%%%%%%%%%%%%%%%%%%%%%%%%%%%%%%%%%%%%%%%%%%%%%%%%%
\section{Application to model distributions}
\label{sec:apply_model}

In this section we apply the tail-regression estimator to synthetic
data.
We construct the seed model distributions as linear combinations of
$H_\mu(A)$, defined in Eq.\ \ref{eq:ptest}, to study various cases of
interest.

%%%%%%%%%%%%%%%%%%%%%%%%%%%%%%%%%%%%%%%%%%%%%%%%%%%%%%%%%%%%%%%%%%%%%%
\subsection{Distribution with undefined fourth moment}
\label{sec:model_mom4}

Distributions with $3<\mu\leq 5$ have convergent standard estimators
for the expectation value and variance, but the uncertainty on the
standard estimator of the variance is divergent.
To exemplify this case we choose to analyze $P_A(A) = \frac 1 2
H_{3.1}(A) + \frac 1 2 H_{4.1}(A)$, which has a leading-order tail
exponent of $\mu=3.1$, close to the lower limit of $3$, and
$\Delta=1$.
The analytical variance of this distribution is $\sigma^2=4.6586$, and
$P_A(A)$ satisfies the analytical limit $y \sim 0.0997 + 0.0730 x$ as
$x \to 0$.

\begin{figure}[htb!]
  \centering
  \includegraphics[width=0.47\textwidth,valign=b]{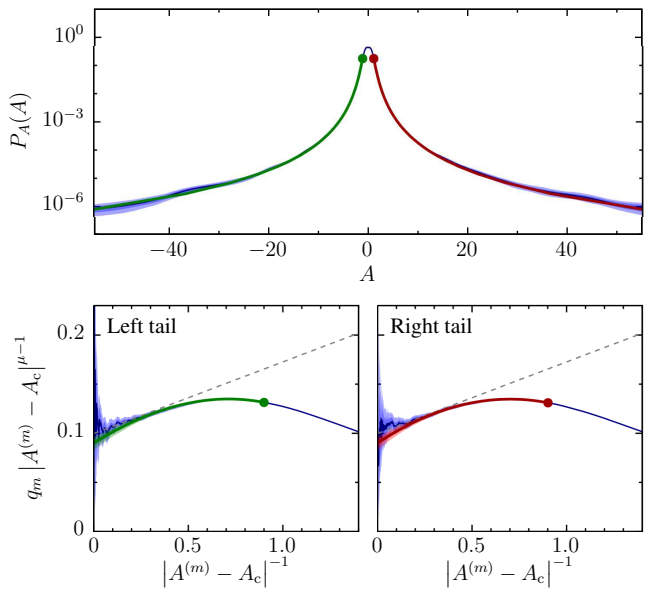}
  \caption{
    Application of the tail-regression procedure to a sample of $10^6$
    random variables distributed according to $P_A(A) = \frac 1 2
    H_{3.1}(A) + \frac 1 2 H_{4.1}(A)$.
    The top panel shows the estimated probability distribution,
    and the lower panels show $yx$ plots for each tail.
    Fits are shown as thick lines in the three panels, and the
    analytical asymptotic form of the tail is shown as a dashed line
    in each of the bottom panels.
    $68.3\%$ and $95.4\%$ confidence intervals obtained from the
    bootstrap are shown as shaded areas.
    \label{fig:h31_h41_fit}}
\end{figure}
The regression of the tails of this model distribution is demonstrated
in Fig.\ \ref{fig:h31_h41_fit} using a sample of $10^6$ random
variables.
The top panel shows the probability distribution estimated by
convolving the data with a variable-width Gaussian kernel, and
the lower panels show $yx$ plots of the data.
The tail fits are shown in the three panels as thick lines, for which
we use $n_{\rm R} = 3$ and $-\log q_{\rm R} = 2.25$, and the
constraint $c_0^{\rm L}=c_0^{\rm R}$ has been imposed.

\begin{figure}[htb!]
  \centering
  \includegraphics[width=0.47\textwidth,valign=b]{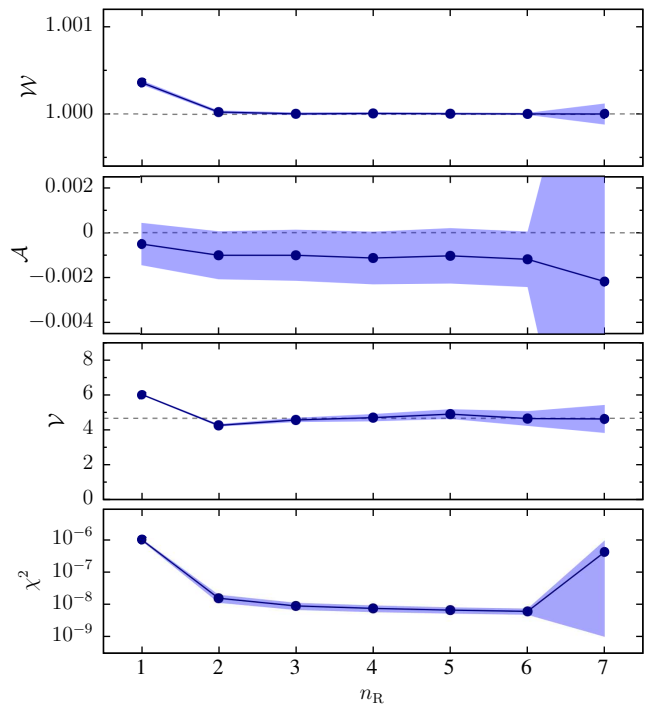}
  \caption{
    Tail-regression estimator of the zeroth, first, and second moment
    of $P_A(A) = \frac 1 2 H_{3.1}(A) + \frac 1 2 H_{4.1}(A)$ and
    $\chi^2$ as a function of expansion order obtained using a sample
    of $10^6$ random numbers and a threshold of $-\log q_{\rm R} =
    2.25$.
    The ``optimal'' $n_{\rm R}=3$ is that at which all of these
    functions reach their respective plateaus.
    \label{fig:h31_h41_assess_n}}
\end{figure}
The process of selecting $n_{\rm R}$ at fixed threshold is illustrated
in Fig.\ \ref{fig:h31_h41_assess_n}, where $\cal W$, $\cal A$, $\cal
V$, and $\chi^2$ are plotted as a function of $n_{\rm R}$.
These functions converge relatively quickly with $n_{\rm R}$, but
for large expansion orders overfitting becomes an issue.
This can be easily spotted by the significant jump in the uncertainty
on the estimators, which in Fig.\ \ref{fig:h31_h41_assess_n} occurs at
$n_{\rm R}=7$.
In this case we consider $n_{\rm R}=3$ to be the ``optimal'' expansion
order.

\begin{figure}[htb!]
  \centering
  \includegraphics[width=0.47\textwidth,valign=b]{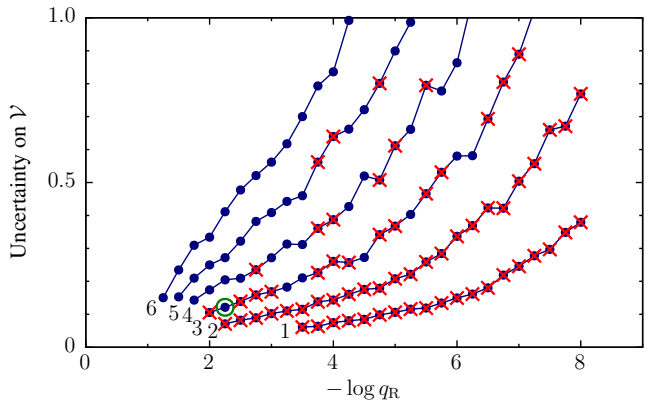}
  \caption{
    Uncertainty on the tail-regression estimator of $\sigma^2$
    obtained from $10^6$ random variables distributed according to
    $P_A(A) = \frac 1 2 H_{3.1}(A) + \frac 1 2 H_{4.1}(A)$
    as a function of threshold $-\log q_{\rm R}$.
    Each curve corresponds to a different expansion order $n_{\rm
    R}$, as labelled.
    Combinations of $-\log q_{\rm R}$ and $n_{\rm R}$ that fail to
    satisfy our correctness criteria are not shown, and narrow fails
    are shown as crossed-out points.
    The ``optimal'' choice of threshold and expansion order, marked
    with a circle, is that which minimizes the uncertainty on $\cal
    V$, corresponding to $-\log q_{\rm R}=2.25$ and $n_{\rm R}=3$ in
    this case.
    \label{fig:h31_h41_assess_anchor}}
\end{figure}
In Fig.\ \ref{fig:h31_h41_assess_anchor} we plot the uncertainty on
$\cal V$ as a function of the threshold for each considered expansion
order.
The uncertainty on $\cal V$ is largely monotonic in $n_{\rm R}$ at
fixed threshold and in $-\log q_{\rm R}$ at fixed expansion order.
Figure \ref{fig:h31_h41_assess_anchor} shows that the ``optimal''
values of the expansion order and threshold are $n_{\rm R} = 3$ and
$-\log q_{\rm R} = 2.25$, as used in Fig.\ \ref{fig:h31_h41_fit}.

\begin{figure}[htb!]
  \centering
  \includegraphics[width=0.47\textwidth,valign=b]{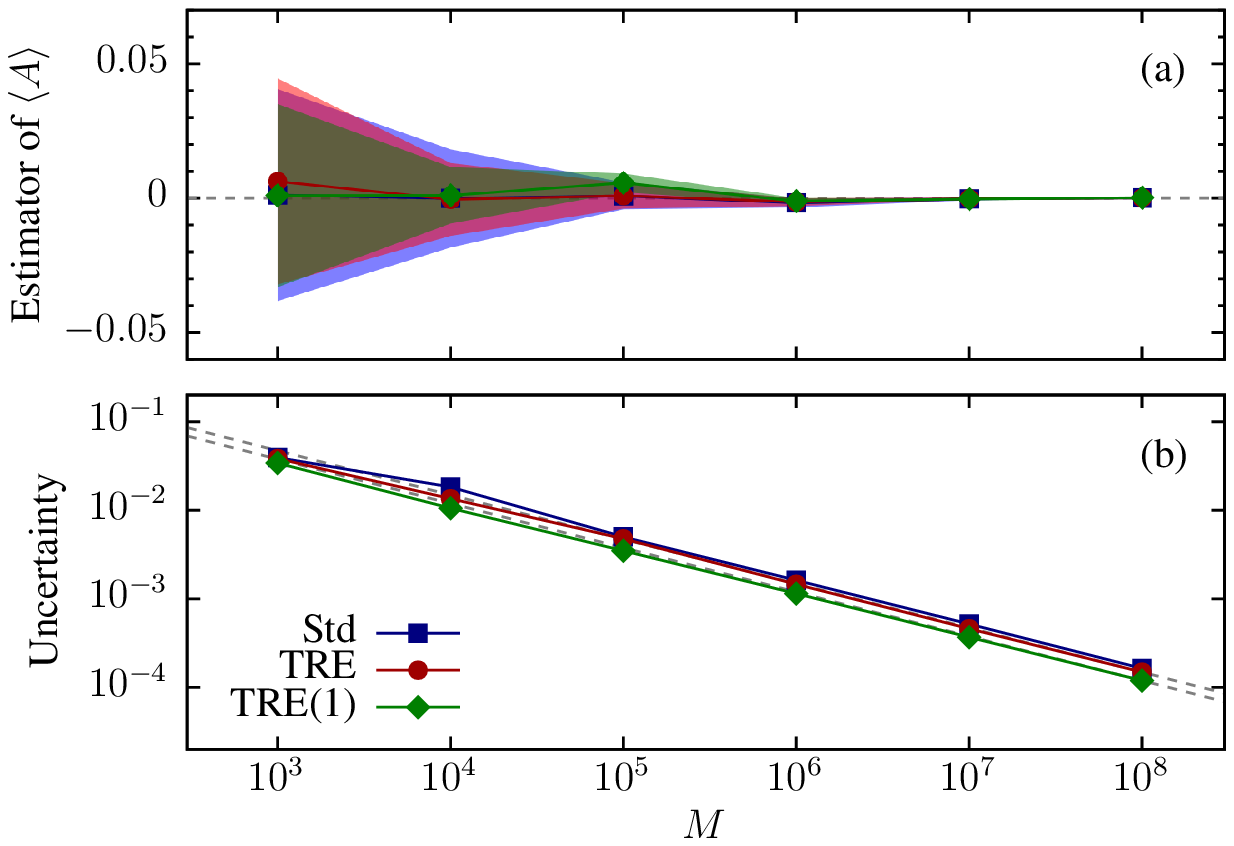}
  \caption{
    Convergence of (a) the estimators of $\langle A\rangle$ and (b)
    their uncertainties as a function of sample size $M$ for model
    distribution $P_A(A) = \frac 1 2 H_{3.1}(A) + \frac 1 2
    H_{4.1}(A)$.
    The exact value $\langle A\rangle=0$ is marked with a dashed line
    in (a), and dashed lines proportional to $M^{-1/2}$ passing
    through the last point of each of the tail-regression estimator
    curves are shown in (b) as guides to the eye.
    $68.3\%$ confidence intervals are shown as shaded areas.
    \label{fig:converge_h31_h41_A}}
\end{figure}

\begin{figure}[htb!]
  \centering
  \includegraphics[width=0.47\textwidth,valign=b]{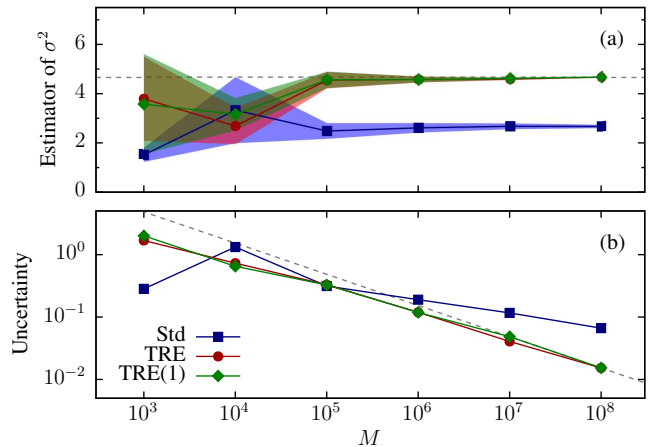}
  \caption{
    Convergence of (a) the estimators of $\sigma^2$ and (b) their
    uncertainties as a function of sample size $M$ for model
    distribution $P_A(A) = \frac 1 2 H_{3.1}(A) + \frac 1 2
    H_{4.1}(A)$.
    The exact value $\sigma^2=4.6586$ is marked with a dashed line in
    (a), and dashed lines proportional to $M^{-1/2}$ passing through
    the last point of each of the tail-regression estimator curves are
    shown in (b) as guides to the eye.
    $68.3\%$ confidence intervals are shown as shaded areas.
    \label{fig:converge_h31_h41}}
\end{figure}

In Table \ref{tab:converge_h31_h41} we report the results obtained
from the standard and tail-regression estimators of $\langle A\rangle$
and $\sigma^2$ for sample sizes ranging from $10^3$ to $10^8$, which
we plot in Figs.\ \ref{fig:converge_h31_h41_A} and
\ref{fig:converge_h31_h41}.
\begin{table*}[ht!]
  \begin{tabular}{ccr@{.}lr@{.}lcccr@{.}lr@{.}lcccr@{.}lr@{.}l}
    \hline
    \multicolumn{1}{c}{                   } & \quad &
    \multicolumn{4}{c}{Standard           } & \quad &
    \multicolumn{6}{c}{TRE (unconstrained)} & \quad &
    \multicolumn{6}{c}{TRE (1 constraint) } \\
    \multicolumn{1}{c}{$M                $} & &
    \multicolumn{2}{c}{$\bar A           $} &
    \multicolumn{2}{c}{$S^2              $} & &
    \multicolumn{1}{c}{$n_{\rm R}        $} &
    \multicolumn{1}{c}{$-\log q_{\rm R}  $} &
    \multicolumn{2}{c}{$\cal A           $} &
    \multicolumn{2}{c}{$\cal V           $} & &
    \multicolumn{1}{c}{$n_{\rm R}        $} &
    \multicolumn{1}{c}{$-\log q_{\rm R}  $} &
    \multicolumn{2}{c}{$\cal A           $} &
    \multicolumn{2}{c}{$\cal V           $} \\
    \hline
    \hline
    $10^3$            & & $ 0$&$001(39)  $ & $1$&$55(28) $ & &
           $1$ & $3.25$ & $ 0$&$006(38)  $ & $3$&$8(17)  $ & &
           $5$ & $1.00$ & $ 0$&$001(34)  $ & $3$&$6(20)  $ \\
    $10^4$            & & $-0$&$000(18)  $ & $3$&$3(13)  $ & &
           $4$ & $1.25$ & $-0$&$000(14)  $ & $2$&$71(73) $ & &
           $3$ & $1.50$ & $ 0$&$001(11)  $ & $3$&$18(65) $ \\
    $10^5$            & & $ 0$&$0009(50) $ & $2$&$51(31) $ & &
           $4$ & $1.50$ & $ 0$&$0009(48) $ & $4$&$54(33) $ & &
           $4$ & $1.50$ & $ 0$&$0057(35) $ & $4$&$55(33) $ \\
    $10^6$            & & $-0$&$0017(16) $ & $2$&$63(19) $ & &
           $3$ & $2.25$ & $-0$&$0014(15) $ & $4$&$57(12) $ & &
           $3$ & $2.25$ & $-0$&$0011(11) $ & $4$&$56(12) $ \\
    $10^7$            & & $-0$&$00022(52)$ & $2$&$69(12) $ & &
           $2$ & $3.75$ & $-0$&$00017(46)$ & $4$&$585(41)$ & &
           $3$ & $2.75$ & $-0$&$00032(37)$ & $4$&$608(49)$ \\
    $10^8$            & & $ 0$&$00018(16)$ & $2$&$689(66)$ & &
           $3$ & $2.75$ & $ 0$&$00016(15)$ & $4$&$661(15)$ & &
           $3$ & $2.75$ & $ 0$&$00019(12)$ & $4$&$662(15)$ \\
    Exact             & & $ 0$&$0        $ & $4$&$6586   $ & &
               &        & $ 0$&$0        $ & $4$&$6586   $ & &
               &        & $ 0$&$0        $ & $4$&$6586   $ \\
    \hline
  \end{tabular}
  \caption{
    Standard and tail-regression estimators of $\langle A \rangle$
    and $\sigma^2$ for model distribution $P_A(A) = \frac 1 2
    H_{3.1}(A) + \frac 1 2 H_{4.1}(A)$ obtained from random samples of
    various sizes $M$.
    The ``optimal'' expansion orders $n_{\rm R}$ and thresholds
    $-\log q_{\rm R}$ used for the tail-regression estimator in
    each case are also shown.
    \label{tab:converge_h31_h41}
  }
\end{table*}
For the tail-regression estimator we report results both without the
use of constraints and imposing $c_0^{\rm L}=c_0^{\rm R}$.
The ``optimal'' values of $n_{\rm R}$ and $-\log q_{\rm R}$ are not
significantly sample-size dependent; we find that $n_{\rm R}$
increases when $-\log q_{\rm R}$ decreases and vice versa, as could be
expected.

The estimators of $\langle A\rangle$ are within uncertainty of the
exact value of zero at all sample sizes, but $\cal A$ has an
uncertainty about $25\%$ smaller than $\bar A$.
The uncertainty on the standard estimator of the variance $S^2$ is
nonmonotonic, as expected, and in this case confidence interval sizes
are severely underestimated, causing the false impression that $S^2$
converges to an incorrect value with increasing $M$.
By contrast, $\cal V$ remains within uncertainty of $\sigma^2$ at all
$M$, and its uncertainty decreases monotonically with sample size.
The unconstrained and constrained estimators give indistinguishable
results in this case, and the uncertainty on both seems to
asymptotically decay as $M^{-1/2}$.

%%%%%%%%%%%%%%%%%%%%%%%%%%%%%%%%%%%%%%%%%%%%%%%%%%%%%%%%%%%%%%%%%%%%%%
\subsection{Distribution with undefined second moment}
\label{sec:model_mom2}

Distributions with $2<\mu\leq 3$ have a divergent variance, and the
standard estimator of $\langle A\rangle$ has an undefined uncertainty.
We exemplify this case with model distribution $P_A(A) = \frac 1 2
H_{2.1}(A) + \frac 1 2 H_{3.1}(A)$, which has $\mu=2.1$, close to
the lower limit of $2$, and $\Delta=1$.

\begin{figure}[htb!]
  \centering
  \includegraphics[width=0.47\textwidth,valign=b]{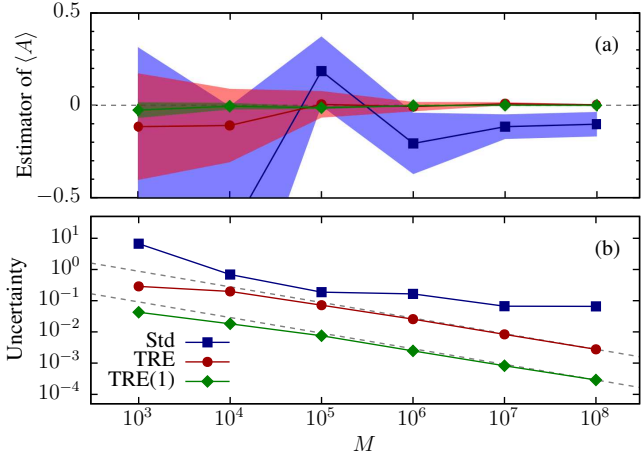}
  \caption{
    Convergence of (a) the estimators of $\langle A \rangle$ and (b)
    their uncertainties as a function of sample size $M$ for model
    distribution $P_A(A) = \frac 1 2 H_{2.1}(A) + \frac 1 2
    H_{3.1}(A)$.
    The exact value $\langle A \rangle=0$ is marked with a dashed line
    in (a), and dashed lines proportional to $M^{-1/2}$ passing
    through the last point of each of the tail-regression estimator
    curves are shown in (b) as guides to the eye.
    $68.3\%$ confidence intervals are shown as shaded areas.
    \label{fig:converge_h21_h31}}
\end{figure}

The standard and tail-regression estimators of $\langle A\rangle$ are
given in Table \ref{tab:converge_h21_h31} and plotted in Fig.\
\ref{fig:converge_h21_h31} as a function of sample size $M$.
\begin{table*}[ht!]
  \begin{tabular}{ccr@{.}lcccr@{.}lcccr@{.}l}
    \hline
    \multicolumn{1}{c}{                   } & \quad &
    \multicolumn{2}{c}{Standard           } & \quad &
    \multicolumn{4}{c}{TRE (unconstrained)} & \quad &
    \multicolumn{4}{c}{TRE (1 constraint) } \\
    \multicolumn{1}{c}{$M                $} & &
    \multicolumn{2}{c}{$\bar A           $} & &
    \multicolumn{1}{c}{$n_{\rm R}        $} &
    \multicolumn{1}{c}{$-\log q_{\rm R}  $} &
    \multicolumn{2}{c}{$\cal A           $} & &
    \multicolumn{1}{c}{$n_{\rm R}        $} &
    \multicolumn{1}{c}{$-\log q_{\rm R}  $} &
    \multicolumn{2}{c}{$\cal A           $} \\
    \hline
    \hline
    $10^3$            & & $-6$&$4(67)    $ & &
           $1$ & $1.30$ & $-0$&$12(29)   $ & &
           $1$ & $1.60$ & $-0$&$026(43)  $ \\
    $10^4$            & & $-0$&$69(69)   $ & &
           $5$ & $1.00$ & $-0$&$11(20)   $ & &
           $4$ & $1.00$ & $-0$&$005(18)  $ \\
    $10^5$            & & $ 0$&$18(19)   $ & &
           $5$ & $1.10$ & $ 0$&$005(72)  $ & &
           $6$ & $1.00$ & $-0$&$0131(76) $ \\
    $10^6$            & & $-0$&$21(17)   $ & &
           $7$ & $1.00$ & $-0$&$007(26)  $ & &
           $4$ & $1.40$ & $-0$&$0021(25) $ \\
    $10^7$            & & $-0$&$115(66)  $ & &
           $5$ & $1.30$ & $ 0$&$0084(84) $ & &
           $4$ & $1.50$ & $ 0$&$00073(83)$ \\
    $10^8$            & & $-0$&$102(66)  $ & &
           $4$ & $1.70$ & $ 0$&$0028(28) $ & &
           $4$ & $1.70$ & $-0$&$00003(29)$ \\
    Exact             & & $ 0$&$0        $ & &
               &        & $ 0$&$0        $ & &
               &        & $ 0$&$0        $ \\
    \hline
  \end{tabular}
  \caption{
    Standard and tail-regression estimators of $\langle A \rangle$
    for model distribution $P_A(A) = \frac 1 2 H_{2.1}(A)
    + \frac 1 2 H_{3.1}(A)$ obtained from random samples of various
    sizes $M$.
    The ``optimal'' expansion orders $n_{\rm R}$ and thresholds
    $-\log q_{\rm R}$ used for the tail-regression estimator in
    each case are also shown.
    \label{tab:converge_h21_h31}
  }
\end{table*}
As expected, the standard estimator $\bar A$ hovers around the exact
value $\langle A\rangle=0$ but its uncertainty does not decrease
uniformly with sample size.
The tail-regression estimator provides a monotonically decreasing
uncertainty with an approximate asymptotic decay proportional to
$M^{-1/2}$, and imposing the constraint $c_0^{\rm L}=c_0^{\rm R}$
yields an order of magnitude smaller uncertainties than the
unconstrained estimator does.

%%%%%%%%%%%%%%%%%%%%%%%%%%%%%%%%%%%%%%%%%%%%%%%%%%%%%%%%%%%%%%%%%%%%%%
\subsection{Symmetric distribution with undefined first moment}
\label{sec:model_mom1}

Distributions with $1<\mu\leq 2$ have a divergent expectation value,
and the standard estimator of $\langle A\rangle$ is undefined.
However if the tails of the distribution are symmetric to leading
order it is possible to redefine $\langle A\rangle$ as a Cauchy
principal value which can be estimated, see Eq.\
\ref{eq:expval_integral_cauchy}.
We exemplify this case with model distribution $P_A(A) = \frac 1 2
H_{1.1}(A) + \frac 1 2 H_{2.1}(A)$, which has $\mu=1.1$, close to
the lower limit of $1$, and $\Delta=1$.

\begin{figure}[htb!]
  \centering
  \includegraphics[width=0.47\textwidth,valign=b]{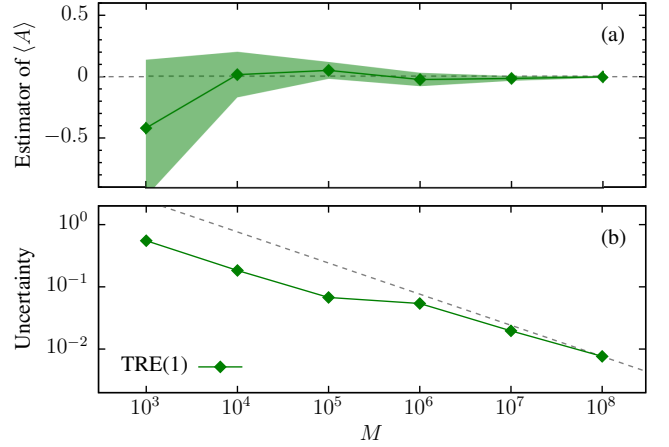}
  \caption{
    Convergence of (a) the constrained tail-regression estimator of
    $\langle A \rangle$ and (b) its uncertainty as a function of
    sample size $M$ for model distribution $P_A(A) = \frac 1 2
    H_{1.1}(A) + \frac 1 2 H_{2.1}(A)$.
    The exact value $\langle A \rangle=0$ is marked with a dashed line
    in (a), and a dashed line proportional to $M^{-1/2}$ passing
    through the last point of the tail-regression estimator
    curve is shown in (b) as a guide to the eye.
    $68.3\%$ confidence intervals are shown as shaded areas.
    \label{fig:converge_h11_h21}}
\end{figure}

Results using the constrained tail-regression estimator are given in
Table \ref{tab:converge_h11_h21} and plotted in Fig.\
\ref{fig:converge_h11_h21}.
\begin{table}[ht!]
  \begin{tabular}{ccccccr@{.}l}
    \hline
    \multicolumn{1}{c}{                  } & \quad &
    \multicolumn{1}{c}{Standard          } & \quad &
    \multicolumn{4}{c}{TRE (1 constraint)} \\
    \multicolumn{1}{c}{$M               $} & \quad &
    \multicolumn{1}{c}{$|\bar A|        $} & \quad &
    \multicolumn{1}{c}{$n_{\rm R}       $} &
    \multicolumn{1}{c}{$-\log q_{\rm R} $} &
    \multicolumn{2}{c}{$\cal A          $} \\
    \hline
    \hline
    $10^3$            & & $\sim 10^{30} $ & &
           $1$ & $1.15$ & $-0$&$42(55)  $ \\
    $10^4$            & & $\sim 10^{28} $ & &
           $2$ & $0.95$ & $ 0$&$01(18)  $ \\
    $10^5$            & & $\sim 10^{39} $ & &
           $2$ & $1.00$ & $ 0$&$046(68) $ \\
    $10^6$            & & $\sim 10^{48} $ & &
           $3$ & $1.10$ & $-0$&$028(54) $ \\
    $10^7$            & & $\sim 10^{59} $ & &
           $6$ & $0.85$ & $-0$&$019(20) $ \\
    $10^8$            & & $\sim 10^{79} $ & &
           $5$ & $0.95$ & $-0$&$0062(76)$ \\
    Exact             & & $ 0.0         $ & &
               &        & $ 0$&$0       $ \\
    \hline
  \end{tabular}
  \caption{
    Standard and tail-regression estimators of $\langle A \rangle$
    for model distribution $P_A(A) = \frac 1 2 H_{1.1}(A)
    + \frac 1 2 H_{2.1}(A)$ obtained from random samples of various
    sizes $M$.
    The ``optimal'' expansion orders $n_{\rm R}$ and thresholds
    $-\log q_{\rm R}$ used for the tail-regression estimator in
    each case are also shown.
    \label{tab:converge_h11_h21}
  }
\end{table}
We find that $\cal A$ is within uncertainty of the exact value of zero
at all sample sizes, and again the uncertainty on $\cal A$ appears to
be asymptotically proportional to $M^{-1/2}$.
In Table \ref{tab:converge_h11_h21} we also give the order of
magnitude of the computed sample mean to highlight that this example
is absolutely intractable with the standard estimator.

%%%%%%%%%%%%%%%%%%%%%%%%%%%%%%%%%%%%%%%%%%%%%%%%%%%%%%%%%%%%%%%%%%%%%%
%%%%%%%%%%%%%%%%%%%%%%%%%%%%%%%%%%%%%%%%%%%%%%%%%%%%%%%%%%%%%%%%%%%%%%
%%%%%%%%%%%%%%%%%%%%%%%%%%%%%%%%%%%%%%%%%%%%%%%%%%%%%%%%%%%%%%%%%%%%%%
\section{Application to variational quantum Monte Carlo data}
\label{sec:apply_vmc}

In this section we explore the performance of the tail-regression
estimator on data obtained from VMC calculations.
Local values of observables generated using the VMC method are usually
affected by serial correlation due to the use of the Metropolis
algorithm to sample configuration space.
In our VMC calculations we perform up to $400$ Metropolis steps
between consecutive evaluations of the local values of the target
observables so that these can be considered to be independent random
variables, and we have verified that the resulting datasets exhibit
negligible serial correlation.

%%%%%%%%%%%%%%%%%%%%%%%%%%%%%%%%%%%%%%%%%%%%%%%%%%%%%%%%%%%%%%%%%%%%%%
\subsection{The energy of the homogeneous electron gas}
\label{sec:ueg}

The homogeneous electron gas is an ideal test bed for methodological
developments in QMC.
We perform VMC calculations on the paramagnetic 54-electron gas in a
cubic simulation cell at density $r_{\rm s} = 1$ using the
Slater-Jastrow (SJ) wave function, consisting of the product of up-
and down-spin Slater determinants of the plane-waves with the smallest
momenta compatible with the periodicity of the simulation cell
multiplied by a Jastrow correlation factor, $\Psi_{\rm SJ}({\bf R}) =
{\rm e}^{J({\bf R})} D_{\uparrow}({\bf R}_{\uparrow})
D_{\downarrow}({\bf R}_{\downarrow})$.
Our Jastrow factor consists of an isotropic electron-electron term of
the Drummond-Towler-Needs form \cite{Drummond_Jastrow_2004,
LopezRios_Jastrow_2012}.
We use the \textsc{casino} code \cite{casino_reference} to generate
samples of local energies whose distribution, as detailed in section
\ref{sec:htail_energy}, has left and right heavy tails of principal
exponent $\mu=4$, $\Delta=1$, and equal left- and right-tail
leading-order coefficients \cite{Trail_htail_2008}.
As explained in section \ref{sec:constraints}, constraints involving
$c_n$ with $n\geq 1$ cannot be applied since $A_0$ is being
approximated by $A_{\rm c}$.

\begin{figure}[htb!]
  \centering
  \includegraphics[width=0.47\textwidth,valign=b]{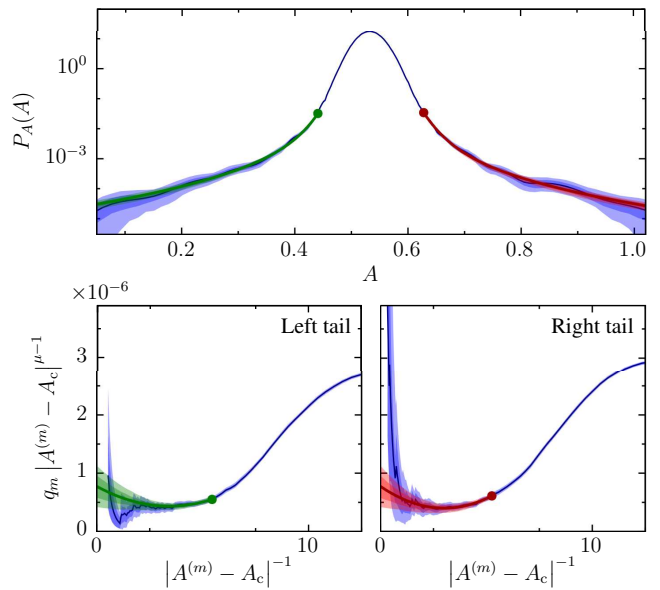}
  \caption{
    Application of the tail-regression procedure to a sample of $10^6$
    VMC local energies for the 54-electron gas at $r_{\rm s}=1$ using
    the SJ wave function.
    The top panel shows the estimated probability distribution,
    and the lower panels show $yx$ plots of both tails.
    Fits are shown as thick lines in the three panels.
    $68.3\%$ and $95.4\%$ confidence intervals obtained from the
    bootstrap are shown as shaded areas.
    \label{fig:ueg_sj}}
\end{figure}

\begin{table*}[htb!]
  \begin{tabular}{ccr@{.}lr@{.}lcccr@{.}lr@{.}lcccr@{.}lr@{.}l}
    \hline
    \multicolumn{1}{c}{                   } & \quad &
    \multicolumn{4}{c}{Standard           } & \quad &
    \multicolumn{6}{c}{TRE (unconstrained)} & \quad &
    \multicolumn{6}{c}{TRE (1 constraint) } \\
    \multicolumn{1}{c}{$M                $} & &
    \multicolumn{2}{c}{$\bar A           $} &
    \multicolumn{2}{c}{$S^2              $} & &
    \multicolumn{1}{c}{$n_{\rm R}        $} &
    \multicolumn{1}{c}{$-\log q_{\rm R}  $} &
    \multicolumn{2}{c}{$\cal A           $} &
    \multicolumn{2}{c}{$\cal V           $} & &
    \multicolumn{1}{c}{$n_{\rm R}        $} &
    \multicolumn{1}{c}{$-\log q_{\rm R}  $} &
    \multicolumn{2}{c}{$\cal A           $} &
    \multicolumn{2}{c}{$\cal V           $} \\
    \hline
    \hline
    $10^3$            & & $0$&$53381(74)  $ & $0$&$000542(34)  $ & &
           $2$ & $4.25$ & $0$&$5339(12)   $ & $0$&$00090(30)   $ & &
           $2$ & $4.50$ & $0$&$53342(96)  $ & $0$&$00095(17)   $ \\
    $10^4$            & & $0$&$53314(24)  $ & $0$&$000581(13)  $ & &
           $2$ & $6.50$ & $0$&$53309(26)  $ & $0$&$000646(40)  $ & &
           $1$ & $7.75$ & $0$&$53310(26)  $ & $0$&$000627(28)  $ \\
    $10^5$            & & $0$&$532868(76) $ & $0$&$0005738(39) $ & &
           $1$ & $8.00$ & $0$&$532863(78) $ & $0$&$0005843(61) $ & &
           $1$ & $8.25$ & $0$&$532868(76) $ & $0$&$0005839(57) $ \\
    $10^6$            & & $0$&$532820(24) $ & $0$&$0005804(44) $ & &
           $2$ & $7.25$ & $0$&$532817(24) $ & $0$&$0005806(24) $ & &
           $2$ & $7.25$ & $0$&$532815(24) $ & $0$&$0005805(24) $ \\
    $10^7$            & & $0$&$5328407(76)$ & $0$&$0005761(15) $ & &
           $2$ & $8.00$ & $0$&$5328396(76)$ & $0$&$00057618(82)$ & &
           $3$ & $8.25$ & $0$&$5328396(76)$ & $0$&$0005763(12) $ \\
    $10^8$            & & $0$&$5328500(24)$ & $0$&$00057442(31)$ & &
           $2$ & $8.25$ & $0$&$5328500(24)$ & $0$&$00057545(25)$ & &
           $2$ & $8.25$ & $0$&$5328499(24)$ & $0$&$00057545(24)$ \\
    \hline
  \end{tabular}
  \caption{
    Standard and tail-regression estimators of $\langle A \rangle$ and
    $\sigma^2$ for the VMC energy of the 54-electron gas at $r_{\rm
    s}=1$ using the SJ wave function obtained from local energy
    samples of various sizes $M$.
    The ``optimal'' expansion order $n_{\rm R}$ and threshold $-\log
    q_{\rm R}$ used for the tail-regression estimator in each case are
    also shown.
    \label{tab:converge_ueg_sj}
  }
\end{table*}
In Fig.\ \ref{fig:ueg_sj} we plot the probability distribution
estimated from $10^6$ local energies, $yx$ plots of the tails, and the
corresponding tail fits.
\begin{figure}[htb!]
  \centering
  \includegraphics[width=0.47\textwidth,valign=b]{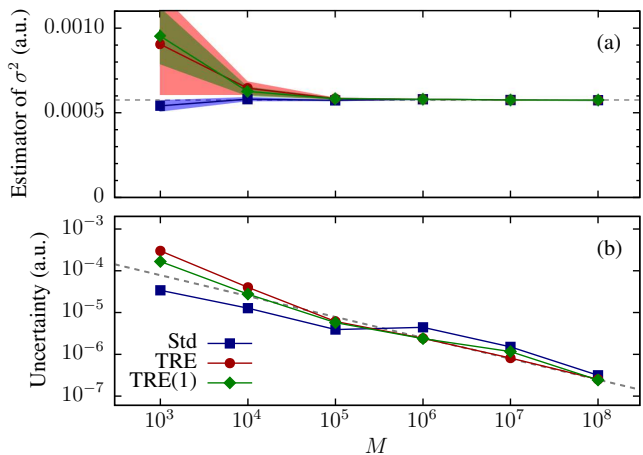}
  \caption{
    Convergence of (a) the estimators of $\sigma^2$ and (b)
    their uncertainties as a function of sample size $M$ for
    the VMC local energy of a 54-electron gas at $r_{\rm s}=1$ using
    the SJ wave function.
    Our best estimate of the value of the variance of the local energy
    for this system, $\sigma^2 \approx 0.00057545(24)$ a.u.\@, is
    marked with a dashed line in (a), and dashed lines proportional to
    $M^{-1/2}$ passing through the last point of each of the
    tail-regression estimator curves are shown in (b) as guides to the
    eye.
    $68.3\%$ confidence intervals are shown as shaded areas.
    \label{fig:converge_ueg_sj}}
\end{figure}
The standard and tail-regression estimators of $\langle A\rangle$ and
$\sigma^2$, given in Table \ref{tab:converge_ueg_sj} and plotted in
Fig.\ \ref{fig:converge_ueg_sj} as a function of sample size, are in
good agreement with each other.
The $\cal A$ estimator offers no advantage over $\bar A$ in this case,
but the uncertainty on $S^2$ exhibits nonmonotonicity as a function
of $M$, while that in $\cal V$ is monotonic, significantly smoother,
and up to 45\% smaller.
Note that even though the nominal standard confidence interval on
$S^2$ only shows minor signs of ill behavior in this example, it is
formally undefined, while the tail-regression estimator produces valid
confidence intervals.
This is of potential practical importance in wave function
optimization and variance extrapolation.

%%%%%%%%%%%%%%%%%%%%%%%%%%%%%%%%%%%%%%%%%%%%%%%%%%%%%%%%%%%%%%%%%%%%%%
\subsection{The atomic force in the C$_2$ molecule}
\label{sec:c2}

We turn our attention to the atomic force in the all-electron carbon
dimer.
The C$_2$ molecule is of particular interest due to its strong
multi-reference character that makes the single-determinantal wave
function incur a large nodal error, which ought to provide relatively
strong heavy tails in the local Pulay and zero-variance force
distributions.

We generate Hartree-Fock orbitals for the all-electron carbon dimer at
an off-equilibrium (compressed) bond length of $r_{\rm CC}=2.0$ a.u.\@
(the experimental equilibrium bond length of C$_2$ is $2.3481$ a.u.\@
\cite{Toulouse_Umrigar_2008}) using the relatively modest cc-pvdz
basis set \cite{Dunning_basis_1992, Kendall_basis_1992} with the
\textsc{molpro} code \cite{molpro}.
We combine these orbitals, modified to satisfy the Kato cusp
conditions at electron-nucleus coalescence points \cite{Ma_cusp_2005},
with a Jastrow factor containing electron-electron, electron-nucleus,
and electron-electron-nucleus terms of the Drummond-Towler-Needs form
\cite{Drummond_Jastrow_2004, LopezRios_Jastrow_2012}, to form the
trial wave function for VMC.
Throughout the VMC run, performed with the \textsc{casino} code
\cite{casino_reference}, we collect local values of the components of
the force on one of the carbon atoms along the molecular axis in the
direction away from the other atom.

\begin{figure}[htb!]
  \centering
  \includegraphics[width=0.47\textwidth,valign=b]{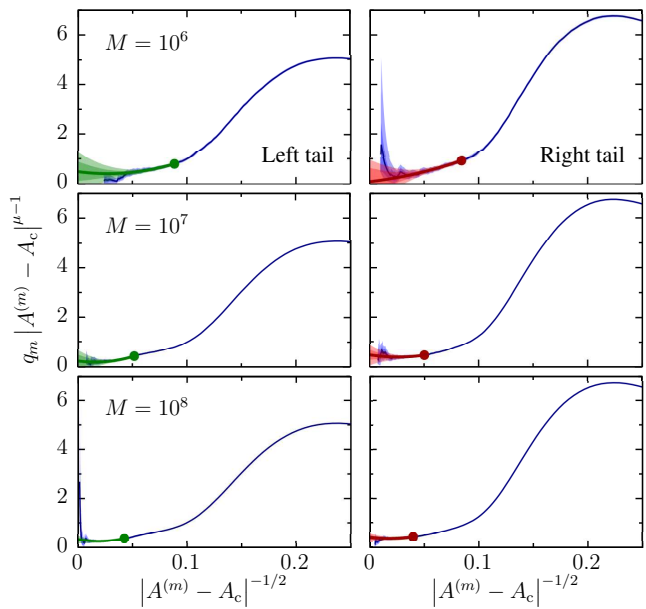}
  \caption{
    $yx$ plots of the left (left) and right (right) tails of the VMC
    local Pulay force on a carbon atom along the molecular axis of the
    C$_2$ molecule at $r_{\rm CC}=2$ a.u.\@ at sample sizes $M=10^6$
    (top), $10^7$ (middle), and $10^8$ (bottom).
    Fits are shown as thick lines in the three panels.
    $68.3\%$ and $95.4\%$ confidence intervals obtained from the
    bootstrap are shown as shaded areas.
    \label{fig:yx_converge_c2_p}}
\end{figure}

We first focus on the Pulay force which, as discussed in Section
\ref{sec:htail_force}, follows a heavy tailed distribution with
$\mu=5/2$ and $\Delta=1/2$ due to the nodal error in the trial wave
function.
In Fig.\ \ref{fig:yx_converge_c2_p} we show $yx$ plots of the tails of
the local Pulay force at sample sizes $M=10^6$, $10^7$, and $10^8$,
along with plots of the corresponding ``optimal'' fits.
Despite having chosen a system known to exhibit a large nodal error,
we find that the leading-order heavy tails are relatively weak, and
that it takes sample sizes of $M \gtrsim 10^7$ to resolve the nonzero
value of $y(0)$.
As a result, the uncertainty on the standard estimator of $\langle
A\rangle$ is likely to only exhibit nonconvergent behavior at large
sample sizes.

\begin{figure}[htb!]
  \centering
  \includegraphics[width=0.47\textwidth,valign=b]{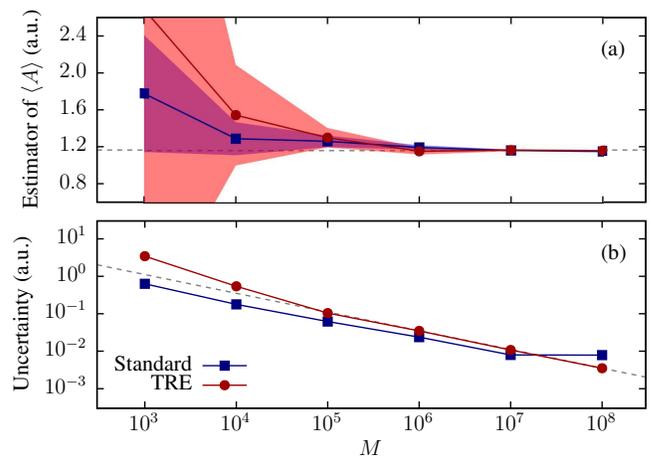}
  \caption{
    Convergence of (a) the estimators of $\langle A\rangle$ and (b)
    their uncertainties as a function of sample size $M$ for the VMC
    local Pulay force on a carbon atom along the molecular axis of the
    C$_2$ molecule at $r_{\rm CC}=2$ a.u.\@.
    The best value of the Pulay force of $1.1643(35)$ a.u.\@ is marked
    with a dashed line in (a), and a dashed line proportional to
    $M^{-1/2}$ passing through the last point of the tail-regression
    estimator curve is shown in (b) as a guide to the eye.
    $68.3\%$ confidence intervals are shown as shaded areas.
    \label{fig:converge_c2_p}}
\end{figure}

We plot the convergence of the standard and tail-regression estimators
of $\langle F_{\rm P}\rangle$ in Fig.\ \ref{fig:converge_c2_p}.
As expected, the standard estimator $\bar A$ seems well behaved at
small sample sizes, but at $M=10^8$ the standard error presents a
substantial nonmonotonic jump.
The uncertainty obtained with the tail-regression estimator remains
smooth and monotonic, and is ultimately smaller than that of the
standard estimator at $M=10^8$.

We find that the local zero-variance corrected Hellmann-Feynman force,
$F_{\rm HFT}+F_{\rm ZV}$, and the local zero-variance corrected total
force, $F_{\rm HFT}+F_{\rm P}+F_{\rm ZV}$, exhibit similarly weak
leading-order tails, and the uncertainties in the standard and
tail-regression estimators follow convergence patterns similar to
those depicted in Fig.\ \ref{fig:converge_c2_p}.

\begin{figure}[htb!]
  \centering
  \includegraphics[width=0.47\textwidth,valign=b]{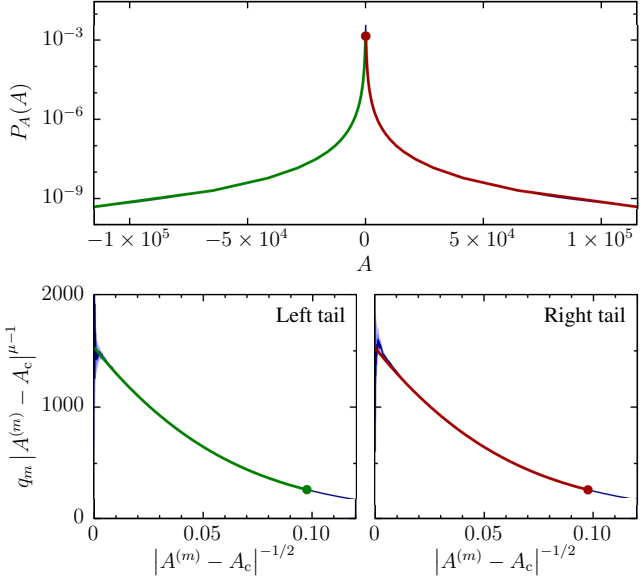}
  \caption{
    Application of the tail-regression procedure to a sample of $10^8$
    VMC local Hellmann-Feynman forces on a carbon atom along the
    molecular axis of the C$_2$ molecule at $r_{\rm CC}=2$ a.u.\@.
    The top panel shows the estimated probability distribution,
    and the lower panels show $yx$ plots of both tails.
    Fits are shown as thick lines in the three panels.
    $68.3\%$ and $95.4\%$ confidence intervals obtained from the
    bootstrap are shown as shaded areas.
    \label{fig:c2_hft_fit}}
\end{figure}

Without the zero-variance correction, the heavy tails affecting the
distribution of the local Hellmann-Feynman force are very strong.
As detailed in Section \ref{sec:htail_force}, these tails are caused
by the presence of all-electron nuclei and the left and right tails
have equal leading-order coefficients.
In Fig.\ \ref{fig:c2_hft_fit} we plot the probability distribution of
$F_{\rm HFT}$ for the off-equilibrium carbon dimer estimated from
$10^8$ sample points and the corresponding $yx$ plots of the left and
right tails of the distribution.
The value of $y(0)$ is very large relative to the rest of the
function, and this is the only case among those we have considered in
which the slope of the $yx$ plot is markedly negative at the origin.
Indeed, it can be shown that the electron-nucleus cusp condition
causes the $c_1$ coefficient in the asymptotic form of $F_{\rm HFT}$
to be approximately proportional to $c_0$ with a large, negative
prefactor.

\begin{figure}[htb!]
  \centering
  \includegraphics[width=0.47\textwidth,valign=b]{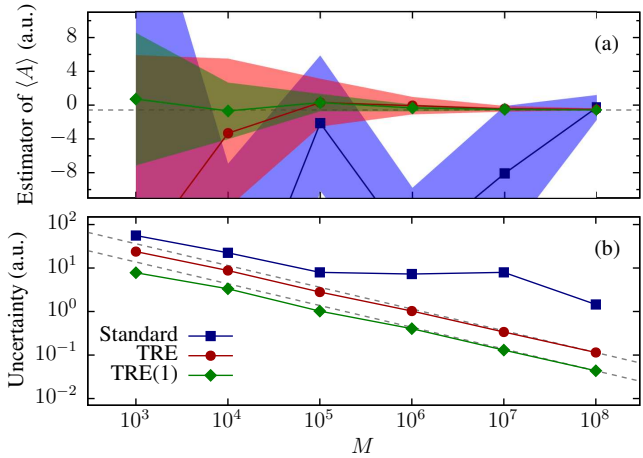}
  \caption{
    Convergence of (a) the estimators of $\langle A \rangle$ and (b)
    their uncertainties as a function of sample size $M$ for the
    VMC local Hellmann-Feynman force on a carbon atom along the
    molecular axis of the C$_2$ molecule at $r_{\rm CC}=2$ a.u.\@.
    The best value of the Hellmann-Feynman force of $-0.5758(29)$
    a.u.\@ is marked with a dashed line in (a), and dashed lines
    proportional to $M^{-1/2}$ passing through the last point of each
    of the tail-regression estimator curves are shown in (b) as guides
    to the eye.
    $68.3\%$ confidence intervals are shown as shaded areas.
    \label{fig:converge_c2_hft}}
\end{figure}

The standard and tail-regression estimators of the expectation value
of the Hellmann-Feynman force are given in Table
\ref{tab:converge_c2_hft} and plotted in Fig.\
\ref{fig:converge_c2_hft} as a function of sample size.
\begin{table*}[ht!]
  \begin{tabular}{ccr@{}lcccr@{}lcccr@{}l}
    \hline
    \multicolumn{1}{c}{                   } & \quad &
    \multicolumn{2}{c}{Standard           } & \quad &
    \multicolumn{4}{c}{TRE (unconstrained)} & \quad &
    \multicolumn{4}{c}{TRE (1 constraint) } \\
    \multicolumn{1}{c}{$M                $} & \quad &
    \multicolumn{2}{c}{$\bar A           $} & \quad &
    \multicolumn{1}{c}{$n_{\rm R}        $} &
    \multicolumn{1}{c}{$-\log q_{\rm R}  $} &
    \multicolumn{2}{c}{$\cal A           $} & \quad &
    \multicolumn{1}{c}{$n_{\rm R}        $} &
    \multicolumn{1}{c}{$-\log q_{\rm R}  $} &
    \multicolumn{2}{c}{$\cal A           $} \\
    \hline
    \hline
    $10^3$            & & $-26$&$(56)     $ & &
           $4$ & $0.90$ & $-18$&$(24)     $ & &
           $3$ & $0.90$ & $  0$&$.7(78)   $ \\
    $10^4$            & & $-29$&$(22)     $ & &
           $5$ & $0.80$ & $ -3$&$.3(88)   $ & &
           $4$ & $0.90$ & $ -0$&$.7(33)   $ \\
    $10^5$            & & $ -2$&$.1(80)   $ & &
           $5$ & $0.80$ & $  0$&$.3(28)   $ & &
           $5$ & $0.80$ & $  0$&$.3(10)   $ \\
    $10^6$            & & $-17$&$.0(73)   $ & &
           $4$ & $1.00$ & $ -0$&$.1(10)   $ & &
           $4$ & $1.00$ & $ -0$&$.34(40)  $ \\
    $10^7$            & & $ -8$&$.0(79)   $ & &
           $3$ & $1.30$ & $ -0$&$.43(34)  $ & &
           $3$ & $1.30$ & $ -0$&$.50(13)  $ \\
    $10^8$            & & $ -0$&$.3(15)   $ & &
           $3$ & $1.40$ & $ -0$&$.51(11)  $ & &
           $3$ & $1.40$ & $ -0$&$.567(43) $ \\
    Best              & & $ -0$&$.5770(29)$ & &
               &        & $ -0$&$.5770(29)$ & &
               &        & $ -0$&$.5770(29)$ \\
    \hline
  \end{tabular}
  \caption{
    Standard and tail-regression estimators of $\langle A \rangle$ for
    the VMC Hellmann-Feynman force on a carbon atom along the
    molecular axis of the C$_2$ molecule at $r_{\rm CC}=2$ a.u.\@,
    obtained from local force samples of various sizes $M$.
    The ``optimal'' expansion orders $n_{\rm R}$ and thresholds
    $-\log q_{\rm R}$ used for the tail-regression estimator in
    each case are also shown.
    The ``best'' value is provided for reference and corresponds to
    the tail-regression estimator of $\langle F_{\rm HFT} +
    F_{\rm ZV}\rangle$ using $10^8$ sample points.
    \label{tab:converge_c2_hft}
  }
\end{table*}
The uncertainty on the standard estimator is clearly nonmonotonic,
while that in the tail-regression estimator is smooth and appears to
present an $M^{-1/2}$ asymptotic decay.
The uncertainty on the tail-regression estimator is up to $60$ times
smaller than the nominal uncertainty on the standard estimator, of
which a factor of $2.5$ is thanks to imposing the analytical
constraint $c_0^{\rm L}=c_0^{\rm R}$.

The results obtained for the different force components at the largest
considered sample size of $M=10^8$ are given in Table
\ref{tab:c2_results}.
\begin{table}[ht!]
  \begin{tabular}{cr@{}lr@{}lr@{}l}
    \hline
    \multicolumn{1}{c}{        } &
    \multicolumn{2}{c}{Standard} &
    \multicolumn{2}{c}{TRE     } &
    \multicolumn{2}{c}{TRE(1)  } \\
    \hline
    \hline
    $\langle F_{\rm HFT} \rangle$                      &
       $ -0$&$.3(15)   $ & $ -0$&$.51(11)  $ & $ -0$&$.567(43) $    \\
    $\langle F_{\rm HFT}+F_{\rm ZV} \rangle$           &
       $ -0$&$.5731(42)$ & $ -0$&$.5770(29)$ & \multicolumn{2}{c}{} \\
    $\langle F_{\rm P} \rangle$                        &
       $  1$&$.1553(78)$ & $  1$&$.1643(35)$ & \multicolumn{2}{c}{} \\
    $\langle F_{\rm HFT}+F_{\rm ZV}+F_{\rm P} \rangle$ &
       $  0$&$.5822(45)$ & $  0$&$.5860(35)$ & \multicolumn{2}{c}{} \\
    \hline
  \end{tabular}
  \caption{
    Standard and tail-regression estimators of $\langle A \rangle$ for
    the VMC local Hellmann-Feynman force, zero-variance corrected
    Hellmann-Feynman force, Pulay force, and zero-variance corrected
    total force on a carbon atom along the molecular axis of the C$_2$
    molecule at $r_{\rm CC}=2$ a.u.\@, obtained from $10^8$ sample
    points.
    \label{tab:c2_results}
  }
\end{table}
The nominal uncertainty on the standard estimator of $\langle F_{\rm
HFT}+F_{\rm ZV}\rangle$ is an order of magnitude smaller than that in
the constrained tail-regression estimator of $\langle F_{\rm
HFT}\rangle$, and in this sense the tail-regression estimator is less
effective than the zero-variance correction.
This is however somewhat misleading since the uncertainty on the
standard estimator remains formally ill-defined, while the
tail-regression estimator is asymptotically normally distributed.
In any case, the tail-regression estimator of $\langle F_{\rm
HFT}+F_{\rm ZV}\rangle$ yields a $30\%$ lower uncertainty than the
standard estimator, which is equivalent to a factor-of-two reduction
in the number of sample points required to achieve a target
uncertainty, showing that the combination of variance-reduction
techniques with the tail-regression estimator is advantageous.
Similarly, the nominal uncertainties on the Pulay force and on the
zero-variance corrected total force are significantly reduced by
replacing the ill-defined standard estimator with the tail-regression
estimator at this sample size.

%%%%%%%%%%%%%%%%%%%%%%%%%%%%%%%%%%%%%%%%%%%%%%%%%%%%%%%%%%%%%%%%%%%%%%
%%%%%%%%%%%%%%%%%%%%%%%%%%%%%%%%%%%%%%%%%%%%%%%%%%%%%%%%%%%%%%%%%%%%%%
%%%%%%%%%%%%%%%%%%%%%%%%%%%%%%%%%%%%%%%%%%%%%%%%%%%%%%%%%%%%%%%%%%%%%%
\section{Application to diffusion quantum Monte Carlo data}
\label{sec:apply_dmc}

At each post-equilibration step of a DMC calculation, an ensemble of
walkers represents the mixed distribution $\Phi({\bf R})\Psi({\bf
R})$, where $\Phi({\bf R})$ is the DMC wave function and $\Psi({\bf
R})$ is the trial wave function.
These walkers carry variable weights which in turn trigger death and
branching events.
The local values $\Psi^{-1}({\bf R}) \hat A \Psi({\bf R})$ of an
observable $\hat A$ are evaluated for each walker at each step, and
the weighted average of the resulting sample yields the mixed
estimator of the expectation value,
\begin{equation}
  \label{eq:mixed_estimator}
  \langle A \rangle =
    \frac{\int \Phi({\bf R}) \hat A \Psi({\bf R}) \, {\rm d}{\bf R}}
         {\int \Phi({\bf R}) \Psi({\bf R}) \, {\rm d}{\bf R}} \;.
\end{equation}
Note that one in principle seeks the pure estimator,
\begin{equation}
  \label{eq:pure_estimator}
  \langle A \rangle =
    \frac{\int \Phi({\bf R}) \hat A \Phi({\bf R}) \, {\rm d}{\bf R}}
         {\int \Phi({\bf R}) \Phi({\bf R}) \, {\rm d}{\bf R}} \;,
\end{equation}
but the mixed estimator is simpler to obtain, it is equal to the pure
estimator if $\hat A$ commutes with the Hamiltonian of the system, and
for other observables there are ways of approximating pure estimators
using mixed estimators \cite{Foulkes_QMC_2001}.
We will restrict our discussion and tests to mixed estimators.

DMC samples differ from VMC samples in important ways.
The formalism presented in section \ref{sec:TRE} can be trivially
altered to accommodate weights, simply by replacing the sample
quantiles $q_m=\frac{m-1/2}M$ with
\begin{equation}
  \label{eq:dmc_sample_quantiles}
  q_m = \frac{\sum_{l: A_l\geq A_m} p_l - p_m/2}
             {\sum_l p_l} \;,
\end{equation}
where $p_m$ is the unnormalized weight of the $m$th sample point.

Walker branching events involve walkers being duplicated and its copies
then evolving independently.
This causes a complex pattern of serial correlation which cannot be
eliminated entirely by leaving several steps between consecutive
evaluations of the local values of observables, as we have done in
our VMC calculations.
While the presence of any form of serial correlation violates our
assumption that samples consist of independent and identically
distributed random variables, we expect this effect to be small
and ignore it in our DMC tests.

The gradient of the DMC wave function $\Phi({\bf R})$ is in general
discontinuous at the nodes \cite{Reynolds_dmc_1981,
Moskowitz_lih_1982}.
This alters the relative presence of walkers on either side of each
nodal point, causing observables whose local values diverge at the
nodes, such as the energy, to exhibit fully asymmetric heavy tails.
The local Hellmann-Feynmann component of the force is not affected by
this, since its singularities do not occur at the nodes of the trial
wave function, so its DMC distribution remains symmetric to leading
order as it is in VMC.

%%%%%%%%%%%%%%%%%%%%%%%%%%%%%%%%%%%%%%%%%%%%%%%%%%%%%%%%%%%%%%%%%%%%%%
\subsection{The atomic force in the C$_2$ molecule}

We have performed a DMC simulation of the C$_2$ molecule at the same
off-equilibrium geometry and with the same wave function as described
in Section \ref{sec:c2}, using a time step of $0.01$ a.u.\@
\cite{Lee_strategies_2011} and a target population of $500$ walkers,
and we have evaluated the local Hellmann-Feynmann and total forces
for each walker every $5000$ steps.

\begin{figure}[htb!]
  \centering
  \includegraphics[width=0.47\textwidth,valign=b]{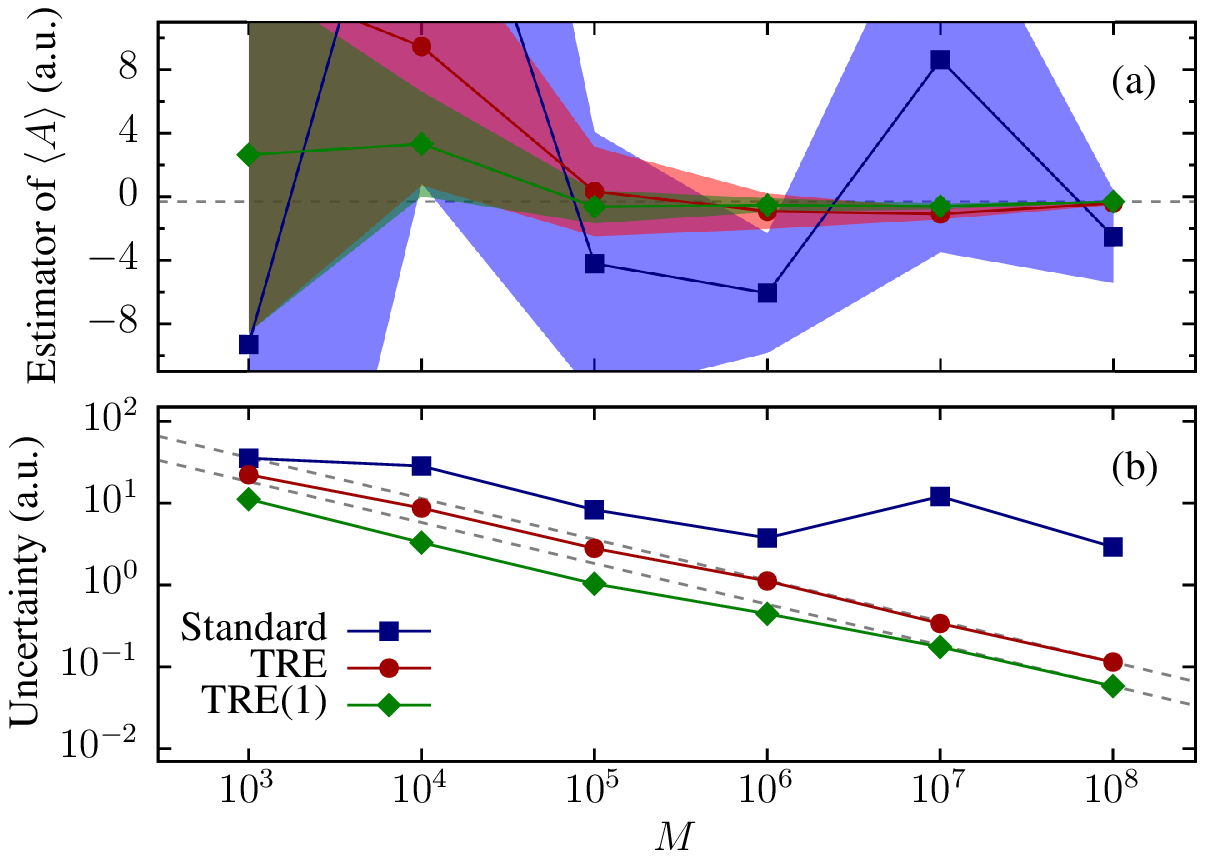}
  \caption{
    Convergence of (a) the estimators of $\langle A \rangle$ and (b)
    their uncertainties as a function of sample size $M$ for the mixed
    DMC local Hellmann-Feynman force on a carbon atom along the
    molecular axis of the C$_2$ molecule at $r_{\rm CC}=2$ a.u.\@.
    The best value of the Hellmann-Feynman force of $0.306(58)$
    a.u.\@ is marked with a dashed line in (a), and dashed lines
    proportional to $M^{-1/2}$ passing through the last point of each
    of the tail-regression estimator curves are shown in (b) as guides
    to the eye.
    $68.3\%$ confidence intervals are shown as shaded areas.
    \label{fig:converge_dmc_c2_hft}}
\end{figure}

The standard and tail-regression estimator of the Hellmann-Feynmann
force are given in Table \ref{tab:converge_dmc_c2_hft} and plotted in
Fig.\ \ref{fig:converge_dmc_c2_hft} as a function of sample size $M$.
These results are very similar to their VMC counterparts; the
nonmonotonic nominal standard error is again up to $60$ times the
uncertainty in the tail-regression estimator, of which a factor of
$2.5$ comes from imposing the constraint $c_0^{\rm L} = c_0^{\rm R}$.
\begin{table*}[ht!]
  \begin{tabular}{ccr@{}lcccr@{}lcccr@{}l}
    \hline
    \multicolumn{1}{c}{                   } & \quad &
    \multicolumn{2}{c}{Standard           } & \quad &
    \multicolumn{4}{c}{TRE (unconstrained)} & \quad &
    \multicolumn{4}{c}{TRE (1 constraint) } \\
    \multicolumn{1}{c}{$M                $} & \quad &
    \multicolumn{2}{c}{$\bar A           $} & \quad &
    \multicolumn{1}{c}{$n_{\rm R}        $} &
    \multicolumn{1}{c}{$-\log q_{\rm R}  $} &
    \multicolumn{2}{c}{$\cal A           $} & \quad &
    \multicolumn{1}{c}{$n_{\rm R}        $} &
    \multicolumn{1}{c}{$-\log q_{\rm R}  $} &
    \multicolumn{2}{c}{$\cal A           $} \\
    \hline
    \hline
    $10^3$            & & $  9$&$(35)    $ & &
           $3$ & $0.90$ & $-14$&$(22)    $ & &
           $3$ & $1.10$ & $ -3$&$(11)    $ \\
    $10^4$            & & $-29$&$(28)    $ & &
           $5$ & $0.80$ & $ -9$&$.5(87)  $ & &
           $4$ & $0.90$ & $ -3$&$.3(33)  $ \\
    $10^5$            & & $  4$&$.2(83)  $ & &
           $4$ & $0.90$ & $ -0$&$.3(28)  $ & &
           $4$ & $0.90$ & $ -0$&$.6(10)  $ \\
    $10^6$            & & $  6$&$.1(38)  $ & &
           $3$ & $1.40$ & $  0$&$.9(11)  $ & &
           $3$ & $1.40$ & $  0$&$.53(45) $ \\
    $10^7$            & & $ -9$&$(12)    $ & &
           $3$ & $1.30$ & $  1$&$.08(34) $ & &
           $4$ & $1.40$ & $  0$&$.61(18) $ \\
    $10^8$            & & $  2$&$.5(29)  $ & &
           $3$ & $1.40$ & $  0$&$.41(11) $ & &
           $4$ & $1.40$ & $  0$&$.306(58)$ \\
    \hline
  \end{tabular}
  \caption{
    Standard and tail-regression estimators of $\langle A
    \rangle$ for the mixed DMC Hellmann-Feynman force on a carbon
    atom along the molecular axis of the C$_2$ molecule at
    $r_{\rm CC}=2$ a.u.\@, obtained from local force samples of
    various sizes $M$.
    The ``optimal'' expansion orders $n_{\rm R}$ and thresholds
    $-\log q_{\rm R}$ used for the tail-regression estimator in
    each case are also shown.
    \label{tab:converge_dmc_c2_hft}
  }
\end{table*}

The results we obtain for the total force at the largest considered
sample size of $M=10^8$ are given in Table \ref{tab:dmc_c2_results}.
In this case, the uncertainty in the tail-regression estimator of the
total force is $10\%$ smaller than the standard error.
\begin{table}[ht!]
  \begin{tabular}{cr@{}lr@{}lr@{}l}
    \hline
    \multicolumn{1}{c}{        } &
    \multicolumn{2}{c}{Standard} &
    \multicolumn{2}{c}{TRE     } &
    \multicolumn{2}{c}{TRE(1)  } \\
    \hline
    \hline
    $\langle F_{\rm HFT} \rangle$                                   &
       $ -2$&$.5(29)   $ & $ -0$&$.41(11)  $ & $-0$&$.306(58)$      \\
    $\langle F_{\rm HFT}+F_{\rm ZV}+F_{\rm P} \rangle$              &
       $  0$&$.5810(42)$ & $  0$&$.5817(38)$ & \multicolumn{2}{c}{} \\
    \hline
  \end{tabular}
  \caption{
    Standard and tail-regression estimators of $\langle A \rangle$ for
    the mixed DMC local Hellmann-Feynman force and total force on a
    carbon atom along the molecular axis of the C$_2$ molecule at
    $r_{\rm CC}=2$ a.u.\@, obtained from $10^8$ sample points.
    \label{tab:dmc_c2_results}}
\end{table}
From these tests we conclude that the tail-regression estimator is
directly applicable to DMC samples generated using relatively long
decorrelation loops, with essentially the same benefits as we have
found for VMC samples.

%%%%%%%%%%%%%%%%%%%%%%%%%%%%%%%%%%%%%%%%%%%%%%%%%%%%%%%%%%%%%%%%%%%%%%
%%%%%%%%%%%%%%%%%%%%%%%%%%%%%%%%%%%%%%%%%%%%%%%%%%%%%%%%%%%%%%%%%%%%%%
%%%%%%%%%%%%%%%%%%%%%%%%%%%%%%%%%%%%%%%%%%%%%%%%%%%%%%%%%%%%%%%%%%%%%%
\section{Conclusions}
\label{sec:conclusions}

We have introduced a conceptually simple estimator of expectation
values for heavy tailed probability distributions whose power-law tail
indices are known.
Unlike the standard estimator, the tail-regression estimator is immune
to the breakdown of the central limit theorem for distributions of
leading-order tail exponent $2<\mu\leq 3$.
Our regression framework is designed to yield asymptotically normally
distributed results, as reflected in the observed asymptotic
$M^{-1/2}$ decay with sample size $M$ of the uncertainty in all of our
tests, and successfully exploits known analytical relations between
leading order tail coefficients to improve the estimation.
We have also demonstrated the estimation of the variance of
distributions of leading-order tail exponent $3<\mu\leq 5$ whose
uncertainty is ill-defined under standard estimation.

Our tests of the tail-regression estimator with variational and
diffusion quantum Monte Carlo data identifies two use cases of
particular practical relevance.
While the standard estimator yields accurate expectation values of the
energy at large enough sample sizes, standard confidence intervals on
the VMC variance of the local energy are formally undefined.
The tail-regression estimator is capable of delivering valid
confidence intervals on the variance which are up to $45\%$ smaller
than those associated with the nominal standard error in our tests.
The tail-regression estimator also yields valid, convergent confidence
intervals on the VMC and DMC atomic force, including the
Hellmann-Feynman force in all-electron systems for which we obtain
uncertainties up to $60$ times smaller than the nominal standard
error.
The combination of the ``zero-variance'' variance-reduction technique
with the tail-regression estimator yields accurate confidence
intervals on the atomic force.

Our present work shows that the principles underpinning the
tail-regression estimator are robust, and systematic use of the
technique for treating quantum Monte Carlo data would be desirable.
However, further work could improve the applicability of our present
formulation.
We have used the bootstrap to enable the evaluation of meaningful
confidence intervals on a range of functions, but this approach should
be replaced with the use of a closed expression for the uncertainty on
the tail-regression estimator in production calculations.
In turn, this would allow the development of an ``on-the-fly''
reformulation of the method that would avoid the need to store all
local values of the desired observables, unaveraged, for later
analysis.
Dropping the requirement that sample points be independent and
serially uncorrelated would also be desirable in order to reduce the
computational cost of the QMC calculations.
With these refinements, the tail-regression estimator will ultimately
represent a great advance in ensuring the statistical soundness of
results obtained from quantum Monte Carlo and similar methods.

%%%%%%%%%%%%%%%%%%%%%%%%%%%%%%%%%%%%%%%%%%%%%%%%%%%%%%%%%%%%%%%%%%%%%%
%%%%%%%%%%%%%%%%%%%%%%%%%%%%%%%%%%%%%%%%%%%%%%%%%%%%%%%%%%%%%%%%%%%%%%
%%%%%%%%%%%%%%%%%%%%%%%%%%%%%%%%%%%%%%%%%%%%%%%%%%%%%%%%%%%%%%%%%%%%%%
\begin{acknowledgments}
  The authors thank John Trail, Neil Drummond, and Richard Needs for
  useful discussions, and acknowledge the financial support of the
  Max-Planck-Gesellschaft, the Engineering and Physical Sciences
  Research Council of the United Kingdom under Grant No.\
  EP/P034616/1, and the Royal Society.
  Supporting research data can be freely accessed at
  \href{https://doi.org/10.17863/CAM.37836}
  {https://doi.org/10.17863/CAM.37836}, in compliance with the
  applicable Open Data policies.
  Our implementation of the tail-regression estimator can also be
  found at \cite{treat_github} and is distributed with the
  \textsc{casino} code \cite{casino_reference}.
\end{acknowledgments}

%%%%%%%%%%%%%%%%%%%%%%%%%%%%%%%%%%%%%%%%%%%%%%%%%%%%%%%%%%%%%%%%%%%%%%
%%%%%%%%%%%%%%%%%%%%%%%%%%%%%%%%%%%%%%%%%%%%%%%%%%%%%%%%%%%%%%%%%%%%%%
%%%%%%%%%%%%%%%%%%%%%%%%%%%%%%%%%%%%%%%%%%%%%%%%%%%%%%%%%%%%%%%%%%%%%%
\appendix
%%%%%%%%%%%%%%%%%%%%%%%%%%%%%%%%%%%%%%%%%%%%%%%%%%%%%%%%%%%%%%%%%%%%%%
%%%%%%%%%%%%%%%%%%%%%%%%%%%%%%%%%%%%%%%%%%%%%%%%%%%%%%%%%%%%%%%%%%%%%%
%%%%%%%%%%%%%%%%%%%%%%%%%%%%%%%%%%%%%%%%%%%%%%%%%%%%%%%%%%%%%%%%%%%%%%
\section{Tail-index estimation methods}
\label{sec:TIE}

Tail-index estimation methods draw inference on the principal exponent
$\mu$ of a power-law heavy tail of leading-order form
$P_A(A)=c_0A^{-\mu}$ at $A\to\infty$.
The Hill estimator \cite{Hill_TIE_1975} of the first-order tail index,
$\mu-1$, is
\begin{equation}
  \label{eq:hill_estimator}
  \frac 1 {\mu-1} \approx
    \frac 1 {M_{\rm R}}
    \sum_{m=1}^{M_{\rm R}} \log A^{(m)} - \log A^{(M_{\rm R}+1)} \;.
\end{equation}
It can be shown that Eq.\ \ref{eq:hill_estimator} is in fact
equivalent to a logarithmic-scale least-squares fit to the tail of the
distribution \cite{Beirlant_TIE_1996}.
Substituting the leading-order form of $P_A(A)$ into Eq.\
\ref{eq:quantile_relation} and taking logarithms yields
\begin{equation}
  \label{eq:beirlant_fit_function}
  \log A^{(m)} \approx
  \frac 1 {\mu-1} \left(-\log q_m \right) +
  \frac 1 {\mu-1} \log\left(\frac{c_0}{\mu-1}\right) \;,
\end{equation}
which is a linear relationship between $\log A^{(m)}$ and
$-\log q_m$ with slope $\frac 1 {\mu-1}$.
Estimation of this slope by linear regression following Eq.\
\ref{eq:beirlant_fit_function} yields Eq.\ \ref{eq:hill_estimator} for
the fitted slope if the $m$th data point is weighted by
\begin{equation}
  \label{eq:beirlant_weight}
  w_m = \left( \log\frac{q_{M_{\rm R}+1}}{q_m} \right)^{-1} \;,
\end{equation}
and $c_0$ is set so that the fit passes through the $(M_{\rm R}+1)$th
point.
The optimal value of $M_{\rm R}$ for the Hill estimator can thus be
found by optimizing a goodness-of-fit measure with respect to $M_{\rm
R}$ \cite{Beirlant_TIE_1996}, such as the $\chi^2$ value of the fit.
Regression methods for tail-index estimation are found to be
particularly robust \cite{Baek_TIE_2010}.

%%%%%%%%%%%%%%%%%%%%%%%%%%%%%%%%%%%%%%%%%%%%%%%%%%%%%%%%%%%%%%%%%%%%%%
%%%%%%%%%%%%%%%%%%%%%%%%%%%%%%%%%%%%%%%%%%%%%%%%%%%%%%%%%%%%%%%%%%%%%%
%%%%%%%%%%%%%%%%%%%%%%%%%%%%%%%%%%%%%%%%%%%%%%%%%%%%%%%%%%%%%%%%%%%%%%

\end{document}